\documentclass{aastex}
\usepackage[onecolumn,numberedappendix]{emulateapj5}

\newcommand{\smyr}{{ M_\odot\ \rm yr^{-1}}}
\newcommand{\sm}{{ M_\odot}}
\newcommand{\cth}{c_{\rm th}}

\newcommand{\tff}{t_{\rm ff}}
\newcommand{\tffs}{t_{{\rm ff},s}}
\newcommand{\tsf}{t_{*f}}
\newcommand{\beq}{\begin{equation}}
\newcommand{\eeq}{\end{equation}}

\newcommand{\ee}{$^{-2}$}
\newcommand{\eee}{$^{-3}$}
\newcommand{\caln}{{\cal N}}
\newcommand{\krho}{{k_\rho}}
\newcommand{\phig}{\phi_{\rm geom}}
\newcommand{\phinon}{\phi_{\rm *non}}
\newcommand{\phipb}{\phi_{{\bar P}}}
\newcommand{\phipc}{\phi_{P,\,\rm core}}
\newcommand{\phirc}{\phi_{\rho,\,\rm core}}
\newcommand{\alv}{\alpha_{\rm vir}}
\newcommand{\mvir}{M_{\rm vir}}
\newcommand{\mg}{M_{\rm g}}
\newcommand{\mbe}{M_{\rm BE}}
\newcommand{\mth}{M_{\rm th}}
\newcommand{\mds}{\dot m_*}
\newcommand{\scl}{\Sigma_{\rm cl}}
\newcommand{\mcl}{M_{\rm cl}}
\newcommand{\psc}{P_{s,\, \rm core}}
\newcommand{\pcl}{P_{\rm cl}}
\newcommand{\rcl}{R_{\rm cl}}
\newcommand{\muh}{\mu_{\rm H}}
\newcommand{\nh}{n_{\rm H}}
\newcommand{\nhref}{n_{\rm H,\,ref}}
\newcommand{\rref}{r_{\rm ref}}

\newcommand{\mcore}{M_{\rm core}}
\newcommand{\msf}{m_{*f}}
\newcommand{\ecore}{\epsilon_{\rm core}}
\newcommand{\rc}{R_{\rm core}}
\newcommand{\fg}{f_{\rm g}}
\newcommand{\lcr}{\langle c^2\rangle}
\newcommand{\lsr}{\langle \sigma^2\rangle}
\newcommand{\lsrcl}{\langle \sigma_{\rm cl}^2\rangle}

\def\lesssim{\mathrel{\hbox{\rlap{\hbox{\lower4pt\hbox{$\sim$}}}\hbox{$<$}}}}
\def\gtrsim{\mathrel{\hbox{\rlap{\hbox{\lower4pt\hbox{$\sim$}}}\hbox{$>$}}}}
\begin{document}
\journalinfo{astro-ph/0206037}
\submitted{Submitted 31st May 2002, Accepted 12th November 2002}
\title{The Formation of Massive Stars from Turbulent Cores}
\author{Christopher F. McKee\altaffilmark{1} and Jonathan C. Tan\altaffilmark{2}}
\affil{1. Departments of Physics \& Astronomy, University of California, Berkeley, CA 94720, USA.}
\affil{2. Princeton University Observatory, Princeton, NJ 08544, USA.}

\slugcomment{Submitted 31st May 2002, Accepted 12th November 2002}

\begin{abstract}
Observations indicate that massive stars in the Galaxy form in regions
of very high surface density, $\Sigma\sim 1$ g cm\ee.  Clusters
containing massive stars and globular clusters have a 
column density comparable to this.  
The total pressure in clouds of such a column density
is $P/k\sim 10^8-10^9$ K cm\eee, far greater than that in the diffuse
interstellar medium or the average in giant molecular clouds.
Observations show that massive star-forming regions are supersonically
turbulent, and we show that the molecular cores out of which
individual massive stars form are as well. 
The protostellar accretion rate in such a core is approximately equal
to the instantaneous mass of the star divided by the free-fall time of
the gas that is accreting onto the star (Stahler, Shu, \& Taam 1980).
The star-formation time 
in this {\it Turbulent Core} model for massive star formation
is several times the mean free-fall time of
the core out of which the star forms, but is about equal to that of
the region in which the core is embedded.  The 
high densities in regions of massive star formation lead to
typical time scales for the formation of a
massive star of about $10^5$~yr.  The corresponding accretion
rate is high enough to overcome the radiation pressure due to the
luminosity of the star.  For the typical case we consider, in which
the cores out of which the stars form have a density structure
$\rho\propto r^{-1.5}$, the protostellar accretion rate grows with
time as $\dot m_*\propto t$.  We present a new calculation of the
evolution of the radius of a protostar and determine the protostellar
accretion luminosity.  At the high accretion rates that are typical in
regions of massive star formation, protostars join the main sequence
at about $20 M_\odot$.
We apply these results to predict the properties of protostars thought
to be powering several observed hot molecular cores, including the
Orion hot core and W3($\rm H_2O$).
In the Appendixes, we discuss the pressure in molecular clouds and we
argue that ``logatropic'' models for molecular clouds are
incompatible with observation.
\end{abstract}

\keywords{hydrodynamics --- ISM: clouds --- stars: formation}

\section{Introduction}

	Massive stars are fundamental in the evolution of galaxies
since they produce the heavy elements, energize the interstellar
medium, and possibly regulate the rate of star formation.  Remarkably
little is known about how massive stars form, however: the problem is
difficult observationally because massive star formation occurs in
distant, highly obscured regions, and it is difficult theoretically
because of the many processes that must be included.  Even such a
basic parameter as the time it takes to form a massive star has
been uncertain.  This time scale, or equivalently, the protostellar
accretion rate, affects the luminosity of the protostar (particularly
for masses $m_*\la 10 M_\odot$ (Palla \& Stahler 1992) and the
strength of protostellar outflows (Richer et al. 2000).  Arguments
based on extrapolating from low-mass star formation lead to formation
times $\tsf>10^6$ yr, a significant fraction of the main-sequence
lifetime of the star (Bernasconi \& Maeder 1996; McLaughlin \& Pudritz
1997, hereafter MP97; Stahler et al. 2000).  Comparison with
observations of hot molecular cores (Osorio, Lizano \& D'Alessio 1999;
Nakano et al. 2000) suggest substantially smaller time scales,
$\tsf\la 10^5$ yr.  An analysis based on observations of protostellar
outflows suggests $\tsf\sim 3\times 10^5$ yr (Behrend \& Maeder 2001).
The small ($\sim 1\times 10^{6}\:{\rm yr}$) spread in ages of stars in
the Orion Nebula Cluster (Palla \& Stahler 1999), where there is no
evidence that the higher mass stars have formed systematically later
compared to the lower-mass population, sets an upper limit of $\tsf\la
1\:{\rm Myr}$ in this case.  What has been lacking is an adequate
understanding of how the formation time is governed by the conditions
in the gas out of which the star forms.

	 Our understanding of low-mass star formation is on a far better
footing, since it has received much more observational and theoretical
attention (Shu, Adams \& Lizano 1987).  Low-mass stars
form by accreting gas from a molecular ``core''
in which gravity overcomes thermal and nonthermal (magnetic
and turbulent) pressure gradients. Shu
(1977) considered the collapse of a singular isothermal sphere, finding
\begin{equation}
\label{eq:shu}
\dot{m}_*=0.975\frac{\cth^3}{G}=4.36\times 10^{-6}
	\left(\frac{T}{20\:{\rm K}}\right)^{3/2}\smyr,
\end{equation}
where $\cth$ is the isothermal sound speed
and the numerical evaluation assumes 
$n_{\rm He}=0.2 n_{{\rm H_2}}$.
Observed temperatures of $10$ to $20\:{\rm K}$ in regions of low-mass
star formation imply accretion rates of about $10^{-6}$ to
$10^{-5}\:{M_\odot\:yr^{-1}}$, consistent with the inferred values
of $\tsf$ for low-mass stars
in these regions 
(Lada 1999).

    There are two difficulties in extending this theory to high-mass
stars.  The first, discussed in some detail 
by Stahler et al. (2000), is that the 
predicted accretion rate depends only on the temperature of the gas.
Once massive stars form, the gas may be heated to 
temperatures $\sim 50$ to $100\:{\rm K}$, but the first massive stars
that form in a region will emerge from gas at $10-20$~K and will have
low accretion rates and correspondingly long formation times.
The second difficulty is feedback from the massive stars.
Since the Kelvin-Helmholtz contraction time is less than the accretion
time for massive stars, they evolve along the main sequence while
accreting.  Massive protostars are thus very luminous, and it
has been suggested that the radiation pressure and ionization they
produce can halt the accretion and determine the upper limit of the
stellar mass function
(Larson \& Starrfield 1971;
Kahn 1974; Wolfire \& Cassinelli 1987; Jijina \& Adams 1996).  
This feedback is so strong that it is impossible
to form stars as massive as those observed if the accretion is assumed
to be spherical, and the discrepancy grows as the accretion rate is
reduced.  These considerations have
motivated the radical suggestion that massive stars form via the coalescence
of low-mass stars in order to achieve a more rapid build
up of the final stellar mass (Bonnell, Bate \& Zinnecker 1998). 

   Recently, McKee \& Tan (2002; hereafter MT) 
addressed the accretion-rate problem.  The first step in
resolving the problem of the apparently low accretion rates
is to realize that, although equation (\ref{eq:shu}) was derived for
an isothermal gas, it should hold approximately when nonthermal support
due to magnetic fields and turbulence is included as well 
(Stahler, Shu \& Taam 1980; Shu et al. 1987).
Observed cores have turbulent motions that increase systematically
with radius (Larson 1981; 
Caselli \& Myers 1995), and this leads to 
an increase in the accretion rate with time
(Myers \& Fuller 1992; Caselli \& Myers 1995; MP97). 
Larger signal speeds allow for the hydrostatic support of denser gas
cores, which then have shorter free-fall times and thus greater
accretion rates once they become unstable. 
We term this the {\it turbulent core} model for massive star formation.

	  The second step in resolving the accretion rate problem is
the recognition that massive stars form in regions of very high
pressure and density.  MT showed that for typical pressures in regions
of massive star formation (Plume et al. 1997), stars form in a time of
order $10^5$~yr.  This result for the time scale is somewhat longer
than that of Osorio et al. (1999), who inferred stellar masses and
accretion rates by comparing calculated spectra with observations.

       The purpose of this paper is several fold. First, we 
determine the relation between the pressure in a molecular cloud and its
surface density $\Sigma$.  
We show that observed regions of massive star formation,
both Galactic and extragalactic, have 
$\Sigma\sim 1$ g cm\ee,
corresponding to mean pressures $\bar P/k\sim 4\times 10^8$ K cm\eee.  
Second, we extend the self-similar theory presented by MT to allow for
magnetic fields and for a thermally supported core.  Finally, we
determine the radius and luminosity of accreting protostars, and
use observed hot core luminosities to predict accretion rates and
masses of several nearby massive protostars.

\section{The Pressure and Surface Density of 
Regions of High-Mass Star Formation}
\label{sec:pressure}

	A cloud in hydrostatic equilibrium has a total pressure
that is proportional to the square of the surface density,
$P\sim G\Sigma^2$, where $\Sigma\equiv M(R)/\pi R^2$ 
(e.g., Elmegreen 1989).  The Appendix contains
a detailed discussion of this relation, and for our fiducial
case---including the effects of magnetic fields and allowing
for embedded stars---we find that the mean
pressure in a cloud is typically
$\bar P\simeq 0.88\, G\Sigma^2$.  
What is the value
of the surface density in regions of current and past massive star
formation?

	Regions of high-mass star formation studied by Plume et al. (1997) are
characterized by virial masses $\mvir\sim 3800\:{ M_\odot}$ and
radii $R\sim 0.5\:{\rm pc}$. 
As discussed in the Appendix, the virial mass $\mvir$ is
related to the actual mass $M$ by the virial parameter, 
$\alv=\mvir/ M$.  The virial parameter is about
$1.3-1.4$ for GMCs, whereas it is quite close to unity for
cores that are actively forming stars.  Although there is no
direct determination of $\alv$ for massive-star forming clumps,
Plume et al. (1997)
regard the virial mass as the most accurate estimate for
the mass $(\alv\simeq 1)$.
Since this is consistent with the results for low-mass star formation
discussed in the Appendix, we shall adopt $\alv=1$ for our numerical
estimates. The mean column density 
in the clumps studied by Plume et al. is then $\scl \simeq
1\:{\rm g\:cm^{-2}}=4800\sm$ pc\ee; the corresponding visual extinction
is $A_V=(N_{\rm H}/2\times 10^{21}$~cm$^{-2})$~mag $= 214\scl$~mag.
From maps of a larger number of sources, Shirley et al (2002) find
a median surface density (which they regard as a better
characterization of their sample) from the virial theorem of
$\scl=0.60$ g cm\ee.  Mueller et al. (2002a,b) have determined the gas
masses of a number of these sources from observations of the 350
\micron\ dust emission.  Using the Ossenkopf and Henning (1994)
opacities, they find gas masses that are on average 3.4 times lower than
the virial mass.  The mean surface density $\scl=0.19\pm 0.12$ g cm\ee\ for
the sources they were able to model in detail, while their total
sample gives a higher value, $\scl=0.73\pm 1.7$ g cm\ee.  Had they
used the Pollack et al. (1994) opacities instead, the inferred
masses and surface densities would have been larger by a factor of
about 2.5.
These data show that there is a considerable dispersion in
the mean surface density of the sources.
These column densities are far greater than those associated
with GMCs (0.035 g cm$^{-2}$---Solomon et al. 1987) or regions
of low mass star formation (the average column density in the 
C$^{18}$O cores in Taurus is 0.032 g cm$^{-2}$---Onishi et al. 1996).

     We now compare the column density observed in regions of
massive star formation with the surface density of stars in both
Galactic and extragalactic star clusters.  Before doing so, we
note that there are several effects that can alter the relation
between the observed stellar surface density and that of the
molecular gas out of which the cluster formed. First, since not
all the gas in the cloud goes into stars, some of the gas is 
ejected, causing the cluster to expand.  For example,
if the star formation were 50\% efficient and the gas were lost slowly,
then the cluster would expand by approximately
a factor of two from its initial size (Hills 1980). 
Including the mass lost from the cluster implies
that the initial surface density of gas would have been 8
times greater than the final surface density of stars.
On the other hand, if magnetic fields were important in the support
of the cloud, then the stellar velocity dispersion would have
been less than the virial velocity (Patel \& Pudritz 1994),
which would lead to an increase in the stellar surface density.
If the size of the cluster is estimated from observations of
the massive stars, then the resulting surface density will
be larger than that of the total star cluster insofar as the massive
stars form preferentially in the inner regions of
clusters (Bonnell \& Davies 1998).
This effect would be amplified if the cluster is old enough
to have experienced significant mass segregation.
Overall, we can expect the stellar surface density to 
be at best within a factor of a few of the initial gas surface density.

	First, consider Galactic star clusters.  
The virial mass of the Orion Nebula Cluster (ONC),
including both stars and gas,
is about $4650 M_\odot$ (averaging
the two models of Hillenbrand \& Hartmann 1998).  
About half the mass of the cluster is within 0.8 pc, 
giving $\Sigma\simeq 0.24$ g cm\ee.  
The Arches is a large cluster near the Galactic Center that
contains many massive stars (Figer et al. 1999).  Kim et al.
(2000) have analyzed this cluster, taking into account the
strong tidal effects experienced by a cluster so near the
Galactic Center.  They conclude that the initial mass of
the cluster was about $2\times 10^4\sm$.  From their
results, we estimate that the
half-mass radius is about 0.4 pc, giving $\Sigma\simeq 4$ g cm\ee. 
Figer et al. infer a very flat IMF for the Arches ($d\caln_*/d\ln m_*
\propto m_*^{-0.6}$ vs. $m_*^{-1.35}$ for the Salpeter IMF).
If the actual IMF of the Arches is closer to the Salpeter
value and the heavy extinction has led to
an underestimate of the number of low-mass stars, then the surface
density would be larger than this estimate.

	Globular clusters in the Galaxy no longer contain any massive
stars, but they presumably did in their youth. There is a considerable
dispersion in the properties of globular clusters; here we estimate
the surface density of a typical one.  The typical mass of a Galactic
globular is about $2\times 10^5 M_\odot$ (inferred from Binney \&
Merrifield 1998). 
The data summarized by van den
Bergh, Morbey, \& Pazder (1991) give a 
median half-light radius for Galactic
globulars of $R_{\rm proj,\, 1/2}=2.6$ pc, where the subscript
``proj" emphasizes that this is a projected radius.
If the light traces the mass, then the 
spherical half-mass radius is $R_{1/2}\simeq 1.31 R_{\rm proj,\,
1/2}$ (Spitzer 1987), which is 3.4 pc.  The surface density inside
the spherical half-mass radius is then $0.5 M/(\pi R_{1/2}^2)
\simeq 0.6$ g cm\eee. 
The corresponding value of the pressure, $\sim 3\times 10^8\:{\rm
K\:cm^{-3}}$, is similar to the estimates of Elmegreen \& Efremov
(1997)
based on the same line of reasoning.

	 It has been suggested that super-star clusters (SSCs) in
external galaxies are globular clusters in the process of formation
(e.g., Ho \& Filippenko 1996).  The SSCs NGC 1569--A1 and A2 each have
masses of about $4\times 10^5 M_\odot$ (Gilbert \& Graham 2001) and
projected half-light radii of about 1.7 pc (DeMarchi et al 1997), corresponding
to $\Sigma\simeq 2.7$ g cm\ee.  A particularly dramatic SSC is the one
observed by Turner, Beck, \& Ho (2000) in NGC 5253: they estimate that
a star cluster with total ionizing luminosity of $\sim 4\times
10^{52}\:{\rm s^{-1}}$ is confined within a region about $1\times 2$
pc in diameter. For zero age main sequence models (Schaerer \& de
Koter 1997), this corresponds to a total stellar mass of 
$0.6-1.5\times
10^6\sm$, for Salpeter extrapolation to 1 and 0.1~$\sm$,
respectively. If we take the projected half-mass radius to be 0.75~pc,
then $\Sigma\simeq 20-50$ g cm\ee.

   These results are collected in Table 1.  It is striking that 
both the molecular regions where massive stars are forming and the 
star clusters that formed from them all 
have surface densities 
$\Sigma\sim 1$ g cm\ee
(with the exception of the cluster in NGC 5253, which is
somewhat larger).
We first characterize the properties of
molecular gas in such an environment and then determine the time scale
for star formation.

\begin{deluxetable}{ccccc} 
\tablecaption{Characteristic Surface Densities of Regions of High-Mass Star Formation\label{tab:sigma}}
\tablewidth{0pt}
\tablehead{
\colhead{Object [Ref.]\tablenotemark{a}} & \colhead{$M$
($\sm$)} & \colhead{$R_{1/2}$ (pc)} & 
\colhead{$\Sigma$ (${\rm g\:cm^{-2}}$)} & \colhead{$\bar \pcl/k$ (${\rm K\:cm^{-3}}$)}
}
\startdata
Galactic Star-forming Clumps [1] & 3800\tablenotemark{b,c} & 0.5\tablenotemark{c} & 1.0 & $4\times10^8$\\
Orion Nebula Cluster [2] & 4600\tablenotemark{b} & 0.8 & 0.24 & $2\times10^7$\\
Arches Cluster [3,4] & $2\times 10^4$ & $0.4$ & $4$ & $7\times10^{9}$ \\
Galactic Globular Clusters [5,6] & $2\times 10^5$ \tablenotemark{b} & 3.4 & 0.8 & $3\times10^8$\\
NGC1569-A1,A2 [7,8] & $4\times 10^5$ \tablenotemark{b} & 2.2 & 2.7 & $3\times10^9$\\
NGC5253 [9] & $0.6-1.5\times 10^6$ \tablenotemark{d} & 1.0 & 20-50 & $2-11\times10^{11}$\\
\enddata
\tablenotetext{a}{\footnotesize References: 
(1) Plume et al. (1997);
(2) Hillenbrand \& Hartmann (1998); 
(3) Figer et al. (1999); 
(4) Kim et al. (2000);
(5) Binney \& Merrifield (1998); (6) van den Bergh et al. (1991); 
(7) Gilbert \& Graham (2001); 
(8) DeMarchi et al. (1997); 
(9) Turner, Beck, \& Ho (2000)}
\tablenotetext{b}{\footnotesize Virial mass estimates}
\tablenotetext{c}{\footnotesize 
The half-mass radius is not well-defined for the
Plume et al. (1997) clouds, since the mass distribution on larger
scales is not known.  We therefore evaluate $\Sigma=M/\pi R^2$ using
the typical radius and virial mass that they observe.}
\tablenotetext{d}{\footnotesize Extrapolation from inferred LyC
luminosity of \ion{H}{2} region based on Salpeter IMF with a lower
mass limit $m_\ell=1,\, 0.1\; M_\odot$.}
\end{deluxetable}

\section{Self-Similar Cores and Clumps}
\label{S:selfsim}

	Molecular clouds are inhomogeneous on a wide range of scales
(Williams, Blitz, \& McKee 2000), 
and numerical simulations show that this is a natural outcome
of supersonic turbulence, both with and without magnetic fields 
(Vazquez-Semadeni et al. 2000).
Following Williams et al., we define a {\it star-forming clump} as a
massive region of molecular gas out of which a star cluster is forming;
a {\it core} is a region of molecular gas that will form a single
star (or multiple star system such as a binary).  Star-forming clumps
are observed to be approximately gravitationally bound (Bertoldi
\& McKee 1992), whereas cores are necessarily gravitationally bound.

	Our basic assumption is that a star-forming clump and the cores
embedded within it are each part of a self-similar, self-gravitating
turbulent structure on all scales above that of the thermal Jeans
mass (MT), an assumption that appears to be in accord with observation
(Williams et al. 2000).  Except on small
scales, the turbulence is therefore necessarily 
supersonic, and we shall show that this is self-consistent below.
We further assume that the structure is approximately stationary in
time (i.e., it is not in a state of dynamical collapse or expansion); 
it follows that 
the star-forming clump and most of the cores within it are
in approximate hydrostatic equilibrium.  However, some of the cores are
gravitationally unstable and are collapsing on a dynamical time
scale---these are the cores that are in the process of forming stars.
Finally, we assume that the clump and the cores within it are
approximately spherical.
As shown by Bertoldi \& McKee (1992), deviations from sphericity of
a factor of a few are readily accounted for; in any case, observations
of high-mass star forming clumps show that they are approximately
spherical (Shirley et al. 2002).

	  This model for
regions of massive star formation is necessarily highly approximate.
Regions of massive star formation are observed to have highly supersonic
velocity dispersions (Plume et al. 1997) and are presumably turbulent.
Large amplitude density fluctuations in such regions continually
form, grow, and dissipate.  Since we have assumed that a clump
has a lifetime that is at least several times greater than its
free-fall time, it follows that only a small fraction of the mass
of the clump can be contained in density fluctuations that are
gravitationally bound and are undergoing gravitational 
collapse---i.e., in cores that are in the process of star formation.
The same conclusion applies to massive cores, which are themselves
turbulent; if on the contrary, most of the mass of a core were
in density fluctuations that were undergoing gravitational collapse,
then the core would form a cluster of low-mass stars instead of
a massive star. 
Calculations of turbulent, self-gravitating clouds indicate
that indeed only a small fraction of the mass is gravitationally
unstable at any given time (e.g., Vazquez-Semadeni, Ballesteros-Paredes,
\& Klessen 2002).

	In this section we shall work out the 
properties of this self-similar structure, leaving consideration of
the effects of thermal pressure to \S \ref{sec:two}.
The results in this section apply equally well to an 
individual core or to a clump.
For clumps, we allow for the possibility that only a fraction
$\fg$ of the mass of the clump is gaseous, with the rest in stars.
If stars are present, we assume that their spatial distribution is 
identical to that of the gas.
We do not attempt to evaluate the mass distribution of the cores,
which is related to the initial mass function (IMF) of the stars that
form; instead, this mass distribution is assumed to be such that it
leads to the observed IMF. Presumably this mass distribution is
determined by the characteristics of the turbulence in the
star-forming clump (e.g., Henriksen \& Turner 1984; Elmegreen \&
Falgarone 1996).

	In a self-similar, spherical 
medium, the density and pressure each have a
power-law dependence on radius,
$\rho\propto r^{-k_\rho}$ and $P\propto r^{-k_P}$.
It follows that the sphere is a polytrope with $P\propto
\rho^{\gamma_p}$. In hydrostatic equilibrium we have (MP96; McKee \& 
Holliman 1999) 
\begin{equation}
\label{eq:k}
k_\rho=\frac{2}{2-\gamma_p},\:\: \:\: \:\: \:\:k_P=\gamma_p k_\rho = 
\frac{2\gamma_p}{2-\gamma_p}=2(k_\rho-1).
\end{equation}
Let $c\equiv (P/\rho)^{1/2}$ be the effective sound speed.
The equation of hydrostatic equilibrium then gives
\begin{equation}
\label{eq:m}
M = \frac{k_P c^2 r}{G},
\end{equation}
where $M$ is the total mass of the system, including any stars
that are present.
It follows that the gas density is
\begin{equation}
\label{eq:rho}
\rho = \frac{A c^2}{2 \pi G r^2},
\end{equation}
where
\beq
A=(3-k_\rho)(k_\rho-1)\fg=\frac{\gamma_p(4-3\gamma_p)\fg}{(2-\gamma_p)^2} 
\label{eq:a}	
\eeq
(cf. MP96).  

	It is convenient to express the properties of the cores
and clumps in terms of the pressure and mass.
Since $\rho=P/c^2$, 
equations (\ref{eq:m}) and (\ref{eq:rho}) give 
\begin{equation}
\label{eq:c}
c=\left(\frac{2\pi G^3 M^2 P}{A k_P^2}\right)^{1/8},
\end{equation}
\begin{equation}
\label{eq:r}
r=\left(\frac{A GM^2}{2\pi k_P^2 P}\right)^{1/4},
\end{equation}
and
\beq
\rho=\left(\frac{Ak_P^2 P^3}{2\pi G^3 M^2}\right)^{1/4}.
\label{eq:rho2}
\eeq
In this subsection, $A$ can refer either to clumps, which generally
have $\fg < 1$, or to cores, which are assumed to have $\fg=1$;
in the remainder of the paper, however, we shall use $A$ to refer
only to cores.

	Observe that the only dependence of these properties on the
polytropic index $\gamma_p$, or equivalently, on the power-law indices
$k_P$ and $k_\rho$, is in the factors $A$ and $k_P$, which are raised
to relatively low powers. Studies of low-mass star formation often
adopt the singular isothermal sphere as a model (Shu 1977), which
appears to be approximately consistent with observation (Andre,
Ward-Thompson, \& Barsony 2000).  Such a model has $\gamma_p=1$, so
that $k_\rho=k_P=2$ and $A=1$.  No data are available on the structure
of cores that are forming very massive stars.  MP96 and MP97 have
discussed an alternative model, the logatrope, in which the pressure
varies as the logarithm of the density, but as discussed in Appendix
B, this model is not physically realistic. Observations of molecular
clouds show that the velocity dispersion $\sigma$ increases outwards
(Larson 1981). Since we expect that $\sigma\propto c$, and since
$c^2=P/\rho \propto r^{2-\krho}$, this implies that $\krho<2$ and
$\gamma_p<1$ (Maloney 1988). (If $\gamma_p$
is written as $\gamma_p=1+1/n$, this corresponds to a negative index
$n$; such a polytrope is therefore referred to as a ``negative index
polytrope.")

For ``high-mass'' cores in Orion, Caselli \& Myers (1995) find
$k_\rho\simeq 1.45$ with a dispersion of $\pm0.2$.  According to van
der Tak et al. (2000), the clumps in which high-mass cores are
embedded have values of $k_\rho$ ranging from 1 to 2, centered around
1.5.\footnote{  
van der Tak et al. (2000) remark that $\krho=1.5$ is just what is expected
for regions undergoing supersonic infall.  In our model, this region
should not extend beyond the core radius---i.e., the radius of
the region that is collapsing to form a massive star.  The core
radius  has a fiducial value $\rc\sim 0.06$ pc (see
eq. \ref{eq:r2}), which corresponds to an angle of only a few arcsec
at the typical distance of several kpc for observed massive
star-forming regions.  The beam sizes used to observe these
regions are generally larger than a few arcsec, so the observed 
values of $\krho$ should
typically refer to the clump in which the core is embedded rather than
to the core itself.
A central assumption of our model is that these
star-forming clumps are in approximate
hydrostatic equilibrium, which implies that the observed values of $\krho$
reflect the structure of this equilibrium, not a dynamical collapse.}
Beuther et al. (2002) find $k_\rho\simeq 1.6\pm 0.5$ for their
sample of massive star-forming regions, while 
Mueller et al. (2002b) find $k_\rho\simeq 1.8\pm 0.4$ for a sample of
31 sources 
(see Evans
et al. 2002 for a review).  For our numerical estimates of the
properties of both high-mass cores and clumps, we shall adopt
$k_\rho=1.5$, which corresponds to $\gamma_p=2/3$, $k_P=1$, and
$A=3/4$.  
For comparison, the value $\krho=1.8$ found by Mueller et al. (2002b)
corresponds to $\gamma_p=8/9$, $k_P=1.6$, and $A=0.96$.

    The value of $\gamma_p$ determines the structure of a polytrope.
As discussed by McKee \& Holliman (1999), 
the stability of the polytrope depends on the adiabatic
index $\gamma$ as well as on $\gamma_p$.  If the polytrope is
isentropic ($\gamma=\gamma_p$), then the maximum density contrast
between the center and edge of a stable molecular cloud is quite limited
($\leq 14$ for $\gamma_p\leq 1$).  On the other hand, for
non-isentropic clouds, an infinite density contrast is possible
in a stable cloud provided $\gamma$ exceeds a critical value;
for our fiducial case of $\gamma_p=2/3$, this critical value
of $\gamma$ is unity.  The value of $\gamma$ for turbulent magnetic fields
is 4/3, significantly above this critical value.  The effective value
of $\gamma$ for the turbulent motions in molecular clouds is
not known; this is a thorny problem, since the motions are
inferred to damp very rapidly 
(Vazquez-Semadeni et al. 2000).  The large density contrasts observed
in molecular clouds suggest that, insofar as polytropic models
are applicable, the effective value of $\gamma$ is large enough
to render the clouds stable against dynamical collapse.

\subsection{The Mean Pressure in Clumps}
\label{sec:meanpr}

	The mean pressure in a clump is $\bar\pcl\equiv(1/V_{\rm
cl}) \int P\, dV$. As discussed in Appendix A, this pressure
is proportional to $G\scl^2$, so
we write
\beq
\bar \pcl\equiv\phipb G\scl^2,
\eeq 
where $\phipb$ is a numerical factor of order unity. From
equation (\ref{eq:pbara}), we find
\beq
\phipb=\left(\frac{3\pi}{20}\right)\fg \phig\phi_B\alv,
\label{eq:phipb}
\eeq
where $\phig$ represents the effect of non-spherical geometry and
$\phi_B\equiv \lcr/\lsr$ the effect of
magnetic fields.  Here $\sigma$ is the one dimensional velocity 
dispersion; in the absence of magnetic fields, this is identical
to the effective sound speed $c$, and $\phi_B=1$.
Insofar as the clump can be approximated as part of a singular
polytropic sphere, equations (\ref{eq:alv2}) and (\ref{eq:alvsps})
show that the product $\phi_B\alv$ is well determined.

	As discussed in the Appendix, clouds with aspect ratios
of 2:1, either oblate or prolate, have $\phig\simeq 1.1$.
Observed low-mass cores have aspect ratios of about 3:1 if they are
oblate and 2:1 if they are prolate (Myers et al. 1991). Basu (2000)
considered triaxial clouds and argued against larger aspect ratios
for oblate clouds, which are formally allowed by the observations.
Although the shape of high-mass cores is unknown, 
high-mass clumps appear to have quite spherical shapes (Shirley et
al. 2002); we therefore 
adopt $\phig=1$ for our fiducial case.
Observations suggest that the Alfven Mach number is of order unity,
and in that case the discussion in the Appendix shows
that $\phi_B\simeq 2.8$. (By contrast, MT considered the non-magnetic
case with $\phi_B=1$.)
The factor $\fg$ measures the fraction of the
cloud mass in gas.  Although the star formation efficiency averaged
over all the molecular gas in the Galaxy is relatively small ($\sim
5\%$ over the life of an association according to Williams \& McKee
1997), we anticipate that the efficiency in the dense clumps forming
massive stars could be considerably higher.
We assume that a typical massive star-forming clump has about
1/3 of its mass in stars, corresponding to $\fg=2/3$.
Finally, as discussed in the Appendix,
the virial parameter $\alv$ is about
unity in star-forming regions.
Adopting these fiducial values for the parameters that enter $\phipb$,
we find that the fiducial value is then
$\phipb\simeq 0.88$,
so that $\bar\pcl=0.88 G\scl^2$ and
\beq
\bar\pcl/k=4.25 \times 10^8\scl^2~~~{\rm K~cm^{-3}}.
\eeq

\subsection{Properties of Cores}
\label{sec:cores}

	We are now in a position to express the properties of a
core out of which a massive star forms in terms of the 
pressure of the clump in which it is embedded and the mass of
the star that will form.  The key point is that, on 
average, the pressure at the surface of a core will
be the same as the ambient pressure in the clump there,
$\psc=\pcl(r)$.  We define the parameter $\phipc$ as the ratio
of this pressure to the mean pressure in the clump.  Since
the clump is self-similar by assumption, we have
\beq
\label{eq:phipc}
\phipc\equiv
\frac{\pcl(r)}{\bar\pcl}=\frac{3-k_P}{3}\left(\frac{r}{\rcl}\right)^{-k_P}
	=\frac{5-2k_\rho}{3}\left[\frac{M(r)}{\mcl}\right]^{-2(k_\rho-1)/
	(3-k_\rho)}.
\eeq
For example, if the clump is a 
singular isothermal sphere ($k_P=k_\rho=2$), 
then $\phipc= \frac 13 [M(r)/\mcl]^{-2}$;
for our fiducial case of $\krho=1.5$, we have $\phipc=\frac 23 [M(r)/\mcl]
^{-(2/3)}.$

What is the typical value of $\phipc$?  If the cores trace the mass of
the clump, then for $k_\rho=1.5$ the value of $\phipc$ at the
half-mass point ($M=0.5\mcl$) is 1.06.  However, Bonnell and Davies
(1998) have argued that, at least in the case of the Orion Nebula
Cluster (ONC), the massive stars formed preferentially in the inner
parts of the cluster.  In particular, the Trapezium stars originated
within about
30\% of the half-mass radius of the ONC.  If we assume that the ONC is
similar to the Plume et al.  clumps and adopt $0.3 R_{1/2}$ as the
typical radius of formation of massive stars, then $\psc\simeq
\pcl(0.3 R_{1/2})$.  For $k_\rho=1.5$, the pressure in the clump
$\propto 1/r$, so $\psc$ for massive star-forming cores is about 3
times that
at the clump surface
(recall that we are identifying $R_{1/2}$ with the radius of the
Plume et al. clumps),
which in turn is a factor $(3-k_P)/3=2/3$ times the mean pressure
inside the clump;
hence, $\phipc\simeq 2$ in this case.  We shall adopt this as a
fiducial value.

The pressure at the surface of the core is then related to the surface
density of the clump by 
\beq
\psc=\phipc\phipb G\scl^2.
\label{eq:psc}
\eeq
Numerically, we have
\beq
\psc/k=4.83\times 10^8\phipc\phipb\scl^2~~~{\rm K~cm^{-3}}
\rightarrow 0.85\times 10^9\scl^2~~~{\rm K~cm^{-3}},
\eeq
where the final expression is for the fiducial case
($\phipc=2.0$ and $\phipb=0.88$).

	To express the core properties in terms of the
final mass of the star that will form out of it, $\msf$, we note
that only a fraction $\ecore$ of the mass of the core will end up in
the star; the rest is dispersed by the powerful
protostellar outflow associated with the star as it forms
(e.g., Matzner \& McKee 2000). We assume that the star formation
efficiency $\ecore$ is constant throughout the accretion process, so
that at any time the mass of a star, $m_*$,
is related to the mass of gas out of
which it formed, $M_{\rm core}$, by
\begin{equation}
\label{eq:eff}
m_{*}=\epsilon_{\rm core}M_{\rm core}.
\end{equation}
The value of $\epsilon_{\rm core}$ is uncertain for massive stars,
since it depends on the unknown properties of protostellar winds from
massive stars. For low-mass stars, $\ecore\sim0.25-0.75$ (Matzner \&
McKee 2000). 
A similar analysis involving the scaling of low-mass
outflows to high-mass protostars stars yields similar efficiencies
(Tan \& McKee 2002a, Paper II), 
so for numerical estimates we shall adopt $\epsilon_{\rm core}=0.5$ as
a typical value.

	We begin with the one-dimensional velocity dispersion
$\sigma=c/\phi_B^{1/2}$.  Since $\phi_B$ is a number of order
unity, $\sigma$ and $c$ scale in the same way with $r$:
\beq
\sigma\propto c\propto r^{(2-k_\rho)/2}\propto
M^{(2-k_\rho)/3(3-k_\rho)}.
\label{eq:sigma}
\eeq
Equation (\ref{eq:c}) implies that the 
velocity dispersion at the surface of a core is given by
\begin{eqnarray}
\label{eq:s2}
\sigma_s=\frac{c_s}{\phi_B^{1/2}}
&=&1.61 \left(\frac{\phipc\phipb}{A k_P^2\phi_B^4\ecore^2}\right)^{1/8}
\left(\frac{m_{*f}}{30 M_\odot}\right)^{1/4}
\scl^{1/4}~~~{\rm km\:s^{-1}},\\
&\rightarrow& 1.27
\left(\frac{m_{*f}}{30 M_\odot}\right)^{1/4}
\scl^{1/4}~~~{\rm km\:s^{-1}},
\end{eqnarray}
where the last value is for the fiducial case.  
This velocity dispersion
is considerably greater than the isothermal sound speed,
$c_{\rm th}=
0.27(T/20$~K$)^{1/2}$ km~s$^{-1}$.  
As a result, the 
cores that form massive stars are necessarily 
supersonically turbulent and therefore clumpy.  
This is one of the principal results of this paper.  It is often
assumed that star formation can occur only when the gas becomes
subsonic (e.g., 
Henriksen 1986; 
Vazquez-Semadeni et al.
2002), and this assumption
leads directly to the problems with massive star formation
that have been cogently summarized by Stahler
et al. (2000).  In our view, this assumption is unjustified,
and we shall show that we can obtain a theory of massive
star formation that appears to be consistent with the data
by dropping it.

	The radius of a core is (eq. \ref{eq:r})	
\begin{eqnarray}
\label{eq:r2}
\rc& =& 0.050
\left(\frac{A}{k_P^2\ecore^2\phipc\phipb}\right)^{1/4}
\left(\frac{m_{*f}}{30 M_\odot}\right)^{1/2}\scl^{-1/2}~~~{\rm pc}\\
&\rightarrow&
0.057\left(\frac{m_{*f}}{30 M_\odot}\right)^{1/2}\scl^{-1/2} 
	~~~{\rm pc}.
\end{eqnarray}
Finally, the density of H atoms at the surface of a core
is (eq. \ref{eq:rho2})
\begin{eqnarray}
\label{eq:n}
n_{{\rm H},s}& =& 1.10\times 10^{6}
\left(A k_P^2\ecore^2\phipc^3\phipb^3\right)^{1/4}
\left(\frac{m_{*f}}{30 M_\odot}\right)^{-1/2}
\scl^{3/2}~~~{\rm cm^{-3}}\\
&\rightarrow& 1.11\times 10^6
\left(\frac{m_{*f}}{30 M_\odot}\right)^{-1/2}
\scl^{3/2}~~~{\rm cm^{-3}}.
\end{eqnarray}

	Note that equations for the clump velocity
dispersion, radius, and density can be obtained from equations
(\ref{eq:s2}), (\ref{eq:r2}), and (\ref{eq:n}) by making the
replacements $\msf/\ecore\rightarrow M_{\rm cl}$ and
$A\rightarrow A_{\rm core}\fg$; by setting
$\phipc=(3-k_P)/3$ so as to set the pressure
equal to its value at the surface of the clump
(eq. \ref{eq:phipc}); and by setting 
$\alv\phi_B=\alpha_{\rm SPS}$
so as to ensure that the clump is in hydrostatic
equilibrium (eq. \ref{eq:alv2}).

	How are the core properties related to those of the clump
in which it is embedded?  
Consider first the surface density
of a core, $\Sigma_{\rm core}\equiv (\msf/\ecore)/\pi \rc^2$.
Equations (\ref{eq:r}) and (\ref{eq:psc}) give
\beq
\frac{\Sigma_{\rm core}}{\scl}=\left(\frac{2k_P^2}{A\pi}\, 
	\phipc\phipb\right)^{1/2}\rightarrow 1.22.
\label{eq:scscl}
\eeq 
The cores thus have a column density comparable to that of the clump
as a whole.
Note that $\scl$ includes the mass of the stars in the clump;
the ratio of $\Sigma_{\rm core}$ to the column density of gas
in the clump is larger by a factor $1/\fg$.

	One of the important questions that must be addressed in any
model for the formation of dense star clusters is how a large number
of protostellar cores can fit into the small volume of the clump in
which the stars are forming.  This issue is automatically solved in
our model since the cores are much denser than the clumps. We compare
the mean density in a core, $\bar\rho_{\rm core}$, to the mean gas
density in the central part of a clump, $\rho_{\rm cl}=\phirc
\bar\rho_{\rm cl}=\phirc 3\fg M/(4\pi R^3)$, where $\bar\rho_{\rm cl}$
is the mean overall clump density and the factor $\phirc=(3-\krho)/3\cdot
(r/\rcl)^{-\krho}$ allows for central concentration. As before we assume the typical
high-mass star-forming core is located at $r=\rcl/3$ (Bonnell \&
Davies 1998), so that for $k_\rho=1.5$, $\phirc\simeq2.6$.
With the aid of equations
(\ref{eq:r}) and (\ref{eq:psc}) we find 
\begin{eqnarray}
\frac{\bar\rho_{\rm core}}{\rho_{\rm cl}}
	&=&\frac{1}{\phirc}\left(\frac{3\phipc}{3-k_P}\right)^{3/4}
	\frac{1}{\fg^{1/4}}
	\left(\frac{\ecore M}{\msf}\right)^{1/2}\\
&\rightarrow& 7.9\left(\frac{\mcl}{4000\; M_\odot}\right)^{1/2}
	\left(\frac{30\;M_\odot}{\msf}\right)^{1/2}.
\label{eq:rhobarratio}
\end{eqnarray}
The mean density of stellar mass cores is about an
order of magnitude larger than the mean gas density in the
surrounding clump.
Thus, even though the clump is forming a very dense star cluster,
there is no ``crowding problem.''

      Finally, we compare the core radius with the tidal radius in
the clump; for our model to be self-consistent, we require the
core radius to be less than the tidal radius.  
If the clump were a rotationally supported disk, then
the tidal radius is 
$R_t\simeq 0.46[\mcore/M(r)]^{1/3}r$, where $r$ is the 
distance from the core to the center of the clump
and $M(r)$ is the clump mass within $r$ (Paczynski 1971). 
For Galactic star-forming clumps, this
is somewhat smaller than the core radius inferred from equation 
(\ref{eq:r}), $\rc=\fg^{-1/4}[\mcore/M(r)]^{1/2}r$.  However,
molecular cloud cores in low-mass star-forming regions
show little evidence for rotation (Goodman et al. 1992), 
and we assume that this
remains true in regions of high-mass star formation.  In that
case, the clump is supported by turbulent motions and magnetic
fields, and the tidal radius is determined by
\beq
(\krho-1)\left[\frac{GM(r)}{r^3}\right]R_t=\frac{G\mcore}{R_t^2},
\eeq
where we have assumed $\krho>1$.
The ratio of the tidal radius to the core radius,
\beq
\frac{R_t}{\rc}=\frac{\fg^{1/4}}{(\krho-1)^{1/3}}\left[\frac{M(r)}
	{\mcore}\right]^{1/6},
\eeq
is then somewhat greater than unity, as required.  Note that this
ratio increases weakly with the clump mass, so that the omission of
effects associated with tidal gravitational fields becomes
a better approximation for more massive star-forming clumps.

\section{Self-Similar High-Mass Star Formation}
\label{sec:self}

      Having determined the properties of massive star-forming clumps
and the cores embedded within them, we now consider the
rate at which a protostar grows once the core from which it
is forming becomes gravitationally unstable.  Dimensional arguments
indicate that the accretion rate should be $\dot m_*\sim \mcore/\tff$, 
where $t_{\rm ff}=(3\pi/32G\rho)^{1/2}$ is the free-fall time 
(Stahler, Shu, \& Taam 1980; see also Shu et al 1987).
Following MT, we write this as
\begin{equation}
\label{eq:dim}
\dot{m}_*=\phi_* \frac{m_*}{t_{\rm ff}},
\end{equation}
where $\phi_*$ is a dimensionless constant of order unity.  This equation has
the same dependence on dimensional parameters as equation
(\ref{eq:shu}) if $\cth$ is replaced by the virial velocity
$(Gm_*/R)^{1/2}$ (Stahler et al 1980;
Shu et al. 1987), since $(Gm_*/R)^{3/2}/G\propto
m_*(Gm_*/R^3)^{1/2}\propto m_*/t_{\rm ff}$.  
More precisely, equations (\ref{eq:m}), (\ref{eq:rho}), and
(\ref{eq:dim})  imply
\begin{equation}
\label{mdot}
\dot{m}_* = \phi_* \epsilon_{\rm core} \frac{4}{\pi \sqrt{3}} k_P
	  A^{1/2} \frac{c^3}{G},
\end{equation}
which is the generalization of equation (\ref{eq:shu}) 
to the non-isothermal case.

      Equation (\ref{eq:dim})
could be violated in the sense that $\phi_*$ is much greater than unity
only in the unlikely case that the star forms from a coherent
spherical implosion; 
if the star formation is triggered by an
external increase in pressure, $\phi_*$ could be increased somewhat, but
deviations from spherical symmetry in the triggering impulse and in
the protostellar core will generally prevent $\phi_*$ from becoming
too large.  The equation could be violated in the opposite sense that
$\phi_*$ is much less than unity if the core is magnetically
dominated, so that collapse could not begin until the magnetic field
diffused out of the core.  However, the magnetic energy 
in star-forming cores is generally 
less than or about equal to 
the gravitational
energy (Crutcher 1999), so this effect is unlikely to be significant.
Note that since the core out of which a massive star forms is highly turbulent,
it will be clumpy and the 
accretion rate can be expected
to show large fluctuations---i.e., in reality $\phi_*$ will be a
random function of time.

       MT showed that if the collapse is
spherical and self-similar, then $\phi_*$ is quite close to unity
provided that the value of $\rho$ entering $t_{\rm ff}$ is evaluated
at the radius in the initial core that just encloses the gas that
goes into the star when its mass is $m_*$.  The resulting
value of $\phi_*$ averages over the fluctuations that would occur
in an actual turbulent core.
For a polytropic sphere, $\rho\propto r^{-k_\rho}$ and $M\propto
r^{3-k_\rho}$, which implies $\rho \propto M^{-k_\rho /
(3-k_\rho)}=M^{2/(4-3\gamma_p)}$. Since we have assumed that
the stellar mass is a constant fraction $\ecore$ of the gas mass
(see eq. \ref{eq:eff}), the accretion rate varies as
\begin{equation}
\label{mdotscaling}
\dot{m}_* \propto m_* \rho^{1/2} \propto m_*^{1-1/(4-3\gamma_p)}.
\end{equation}
Integration yields
\begin{equation}
\label{mstar}
m_* = m_{*f} \left(\frac{t}{t_{*f}}\right)^{4-3\gamma_p},
\end{equation}
where $m_{*f}$ is the final stellar mass, which is attained at a time
$t_{*f}$ (MP97).  
This equation implies
\begin{equation}
\label{mdotMP}
\dot{m}_* = (4-3\gamma_p)\frac{m_*}{t} = (4-3\gamma_p)\frac{m_{*f}}{t_{*f}}
	  \left(\frac{t}{t_{*f}}\right)^{3-3\gamma_p}.
\end{equation}
Note that for $\gamma_p<1$ the accretion rate
accelerates (MP97); in particular,   
for the fiducial case $\gamma_p=2/3$, we have $\mds\propto t$.  From
equations (\ref{eq:dim}) and (\ref{mdotMP}), the star-formation time,
which is the time from the first formation of 
a protostellar
core to the time the star reaches its final mass, is then
\begin{equation}
\label{tsf}
t_{*f}=\frac{(4-3\gamma_p)}{\phi_*} \tffs,
\end{equation}
where $\tffs$ is the value of the free-fall time evaluated at the
surface of the core---i.e., for the last parcel of gas to accrete onto
the star.

As discussed by MP97, termination of the accretion breaks the
self-similarity once the expansion wave reaches $m_{*f}$. This occurs
at a time they label $t_{\rm ew}$, which is in
the range $(0.42-0.49) t_{*f}$ for $0\leq\gamma_p\leq 1$.
If the boundary of the core is at constant pressure, a
compression wave will be launched inward when the expansion wave
reaches the boundary (Stahler et al. 1980).
Equation (\ref{mdotMP}) remains accurate until this
compression wave reaches the center. 
Numerical calculations by Reid, Pudritz,
\& Wadsley (2002) show that in the logatropic case 
(which is related to the polytropic case with $\gamma_p\rightarrow 0$)
this occurs at a time of about $0.7\tsf$.
Thereafter, the accretion rate gradually decreases.

\begin{figure*}
\epsscale{1.0}
\plotone{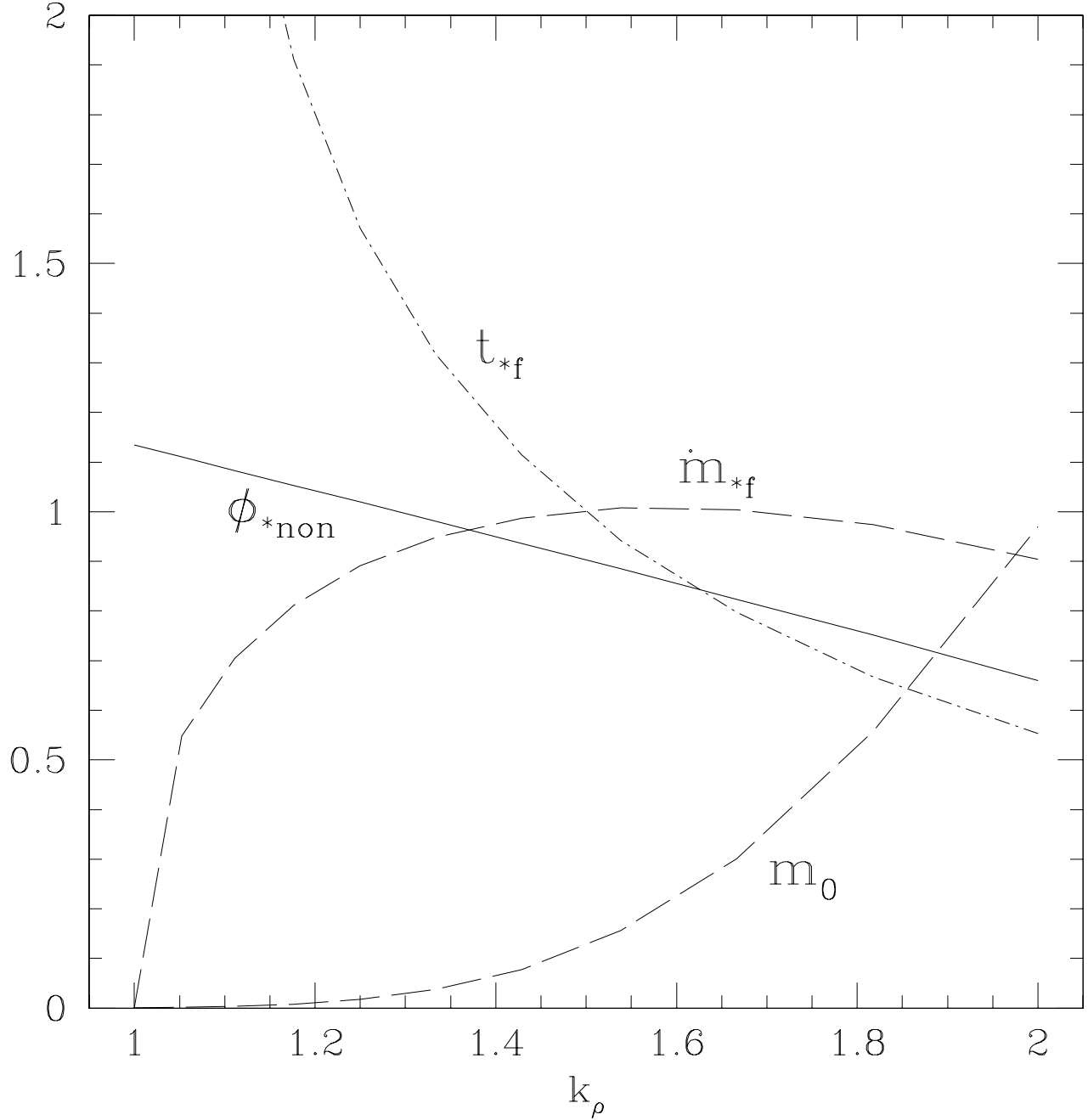}
\caption{Variation of model parameters ($m_0$, $\phinon$) and results
($\dot{m}_{*f}$, $t_{*f}$) with $k_\rho$. Values of $m_0$ are from
Table 4 of MP97.  Over the entire range of $\gamma_p$ and $k_\rho$
relevant to molecular clouds ($0\leq\gamma_p\leq1$, $1\leq k_\rho\leq
2$), $\phinon\simeq 1.62-0.48 k_\rho$ to within about 1\%.  The star
formation time decreases from $3.5t_{\rm ff}$ to $1.5t_{\rm ff}$ as
$\gamma_p$ varies from 0 to 1.  The variation of $\dot{m}_{*f}$ and
$t_{*f}$ relative to the $k_\rho=1.5$ case is also shown.  Note that
the singular polytropic model in hydrostatic equilibrium breaks down
for $k_\rho =1$, $\gamma_p=0$, since then the pressure gradient
vanishes ($k_P=0$).
\label{fig:mzero}}
\end{figure*}

MT used the results of MP97 to evaluate $\phi_*$ in the non-magnetic
case, which we label $\phinon$:
\begin{mathletters}
\label{phi}
\begin{eqnarray}
\label{phi-a}
\phinon & = &(4-3\gamma_p)\frac{\tffs}{t_{*f}},\\
   & = & \pi\surd 3 \left[\frac{(2-\gamma_p)^2(4-3\gamma_p)^
      {(7-6\gamma_p)/2}m_0}{8^{(5-3\gamma_p)/2}}\right]^{1/(4-3\gamma_p)},
\label{phi-b}
\end{eqnarray}
\end{mathletters}
(see Fig. \ref{fig:mzero}),
where $m_0(\gamma_p)$ is a parameter
that is evaluated by MP97.\footnote{MP97 assumed that the gas was
isentropic---i.e., that the adiabatic
index $\gamma$ is the same as the polytropic index $\gamma_p$.  As
discussed above, the large observed density contrasts in molecular
clouds suggest that $\gamma>\gamma_p$.  No calculations are available
for this case, however, so we have adopted the isentropic results of 
MP97.}
For example, for the singular isothermal
sphere, we have $m_0=0.975$ and $\phinon=0.975 \pi \sqrt{3}/8=0.663$.  For
other values of $\gamma_p$ in the range $0\leq\gamma_p\leq 1$, which
is equivalent to $1\leq k_\rho\leq 2$, the approximation 
\beq
\phinon\simeq 1.62-0.48 k_\rho
\eeq
is accurate to within about 1\% (MT).
For the fiducial case ($k_\rho=1.5$), we have $\phinon=0.90$.
We conclude that $\dot m_*\simeq m_*/t_{\rm ff}$ to within a factor
1.5 for non-magnetic,
spherical cores in which the effective sound speed increases
outward.

      We can estimate the effect of magnetic fields on
the accretion rate from the work of Li \& Shu (1997), who 
considered collapse of self-similar, isothermal,
magnetized, toroidal clouds.  The equilibrium surface density is
$\Sigma=(1+H_0)\cth^2/(\pi G \varpi)$, where $\varpi$ is the cylindrical
radius and $H_0$ is a parameter that increases from zero as the
magnetic field is increased.  They show that the accretion rate is
$\dot m_*=1.05(1+H_0)\cth^3/G$, which is 
larger than that in equation (\ref{eq:shu}) by about a factor $(1+H_0)$.
However, equation (\ref{eq:dim}) predicts $\dot m_*\propto
M\rho^{1/2}\propto M^{3/2}/\varpi^{3/2}\propto \Sigma^{3/2}\varpi^{3/2}
\propto (1+H_0)^{3/2}$.
To reconcile this result with the correct value,
we require 
\beq
\phi_*\simeq \frac{\phinon}{(1+H_0)^{1/2}}.
\label{eq:phistar}
\eeq
Li \& Shu regard $H_0\sim 1$ as
typical for low-mass star formation, which would lead to a reduction
in the accretion rate by a factor of about 1.4.

   The star formation time $\tsf$ increases from $1.5(1+H_0)^{1/2}\tffs$ to 
$3.5(1+H_0)^{1/2}\tffs$ as $\gamma_p$ decreases from 1 to 0;
for the fiducial case, $\tsf=2.22(1+H_0)^{1/2}\tffs$.
(Note, however,
that models with $\gamma_p$ close to zero are not realistic because
they have no pressure gradient: $k_P\rightarrow 0$ from eq. \ref{eq:k}.
Also, see the discussion of logatropes in Appendix B).
The ratio of the star formation time to the mean free-fall time
of the core, $\bar t_{\rm ff, \, core}\equiv(3\pi/32 G\bar\rho_{\rm
core})^{1/2}$, is larger by a factor $(\bar \rho_{\rm core}/\rho_s)^{1/2}
=[3/(3-\krho)]^{1/2}$.  For example, for $k_\rho=1.5$,
we have $\tsf=3.14(1+H_0)^{1/2}\bar t_{\rm ff,\, core}$.  Finally,
we note that if we use equation (\ref{eq:rhobarratio}) to
normalize the star formation time to the mean free-fall
time of the clump in which the core is embedded, we find
\beq
\label{tsftffcl}
\tsf=0.98\left(\frac{1+H_0}{2}\right)^{1/2}\left(\frac{\msf}{30\, M_\odot}\right)^{1/4}
\left(\frac{4000 M_\odot}{\mcl}\right)^{1/4} \bar t_{\rm ff,\, cl}.
\eeq  
Hence,
for typical conditions in Galactic regions of high-mass star
formation, the time to form a star is about
equal to the mean free-fall time of the parent clump,
$\tsf\simeq t_{\rm ff,\, cl}$.
In super star clusters, which have $\mcl\sim 10^6\;\sm$,
the star formation time is somewhat
less than this.

\begin{deluxetable}{ccc} 
\tablecaption{Model parameters and their fiducial values\label{tab:param}}
\tablewidth{0pt}
\tablehead{
\colhead{Parameter} & \colhead{Definition} & \colhead{Fiducial Value}
}
\startdata
$A$ & $(3-k_\rho)(k_\rho-1)f_g$ & 1/2 (clumps), 3/4 (cores)\\
$f_{\rm acc}$ & fraction of core's gravitational energy radiated by star+disk & 0.5\\
$f_g$ & gas mass fraction:
$M_{\rm g}/M=\Sigma_{\rm g}/\Sigma$
& 2/3 (clumps), 1 (cores)\\
$H_0$ & magnetic field parameter in Li-Shu cores & 1.0\\
$k_\rho$ & clump and core density structure: $\rho\propto r^{-k_\rho}$ & 3/2\\
$k_P$ & clump and core pressure structure: $P\propto r^{-k_P}$ & $2(k_\rho -1)=1$\\
$m_A$ & Alfven Mach number & 1.0\\
$\alv$ & virial parameter, $5\lsr R / (GM)$ & 1.0\\
$\ecore$ & core star formation efficiency & 0.5\\
$\gamma_p$ & polytropic index: $P\propto \rho^{\gamma_p}$ & $k_P/k_\rho=2/3$\\
$\phig$ & deviation from sphericity: $R_m^3/(R^2Z)$ & 1.0\\
$\phi_B$ & $\lcr/\lsr = 1.3 + 3/(2m_A^2)$ & 2.8\\
$\phipb$ & $\bar{\pcl}/(G\scl^2)=(3\pi/20)f_g\phig\phi_B\alv$  & 0.88\\
$\phipc$ & $\pcl(r)/\bar{\pcl}$  & 2.0\\
$\phirc$ & $\rho_{\rm cl}(r)/\bar{\rho}_{\rm cl}$ & 2.6\\
$\phinon$ & $\dot{m}_* \tff / m_*$ 
~~(non-magnetic case) 
& 0.90\\
$\phi_*$ & $\phinon 
/(1+H_0)^{1/2}$ & 0.64\\
\enddata
\end{deluxetable}

\subsection{Evaluation of the
Protostellar Accretion Rate and Star Formation Timescale}

     The general expression for the accretion rate in terms of the
surface pressure and the final stellar mass can now be inferred from
equations (\ref{mstar})---(\ref{tsf}): 
\beq
\label{eq:mdot}
\dot{m}_* = \phi_*\;\frac{\msf}{\tffs}\left(\frac{m_*}{\msf}\right)^j,
\eeq
where 
\beq
j\equiv \frac{3(2-k_\rho)}{2(3-k_\rho)}. 
\label{eq:j}
\eeq
In the fiducial case
($k_\rho=1.5$), we have $j=0.5$;
for $\krho=1.8$, we have $j=0.25$.
Using equation 
(\ref{eq:rho2}) 
to evaluate $\tffs\propto (1/\rho_{{\rm core},s})^{1/2}$,
we find
\begin{eqnarray}
\label{mdothighmass}
\mds & = & 9.53\times10^{-4} \phi_* A^{1/8} k_P^{1/4}\ecore^{1/4}
\left(\frac{m_{*f}}{30\:{M_\odot}} \right)^{3/4} 
\left(\frac{\psc/k}{10^9\:{\rm K\:cm^{-3}}}\right)^{3/8} 
\left(\frac{m_*}{m_{*f}}\right)^j
\:{M_\odot\: {\rm yr}^{-1}},\\
\label{mdotscl}
& \rightarrow & 4.6\times 10^{-4}
\left(\frac{m_{*f}}{30\:{M_\odot}}\right)
^{3/4} \scl^{3/4} 
\left(\frac{m_*}{m_{*f}}\right)^{0.5}~{M_\odot\:{\rm yr}^{-1}},
\end{eqnarray}
where again 
we have used the symbol $\rightarrow$ to emphasize that
the final 
expression applies only in the fiducial case.
Although the exponent $j$ in the relation $\mds\propto m_*^j$ changes
with $k_\rho$, the numerical coefficient is constant to within 10\% for
$1.3<k_\rho<2$.  The evolution
of the mass-accretion rate and the mass is illustrated in Figure
\ref{fig:mdott}.
Observe that our expression for the accretion rate depends on
two dimensional parameters, the final
stellar mass, $\msf$, and the surface density of the clump, $\scl$
(or, equivalently, the pressure in the clump), as well as
on the ratio of the current stellar mass, $m_*$, to the final.
We have expressed our results in this form 
in order to facilitate comparison with observation.
In fact, the accretion rate depends on only two parameters,
the enclosed mass, $M$,
and the parameter
$K_p\equiv P/\rho^{\gamma_p}$, which is constant for 
polytropes (Tan \& McKee in preparation).

	The fiducial case includes the effect of a magnetic field.
How does the field affect the accretion rate?  
From equations (\ref{eq:phipb}), (\ref{eq:psc}), and (\ref{eq:mdot}) 
we have
\beq
\mds \propto \phi_*\psc^{3/8}\propto \phi_*\phipb^{3/8}
	     \propto \frac{\phi_B^{3/8}}{(1+H_0)^{1/2}}
	     \rightarrow 1.04,
\eeq
since $\phi_B=2.8$ and  $H_0=1$ in the fiducial case.
There is therefore very little difference between
this case and the field-free case considered by MT.  It should
be kept in mind that we have assumed that the value of the virial
parameter $\alv$ has been taken from observation and is therefore fixed
as the field is varied.  As discussed in Appendix A, if the
value of the virial parameter is taken from theory, then
$\alv\phi_B$ is fixed and the accretion rate would vary as
$(1+H_0)^{-1/2}$. 

     The corresponding value of the star-formation time
is
\begin{eqnarray}
\label{tsfhighmass}
t_{*f} & = & 3.15 \times 10^4 
\left(\frac{4-3\gamma_p}{\phi_* A^{1/8} k_P^{1/4}\ecore^{1/4}}\right)
\left(\frac{m_{*f}}{30\:{M_\odot}} \right)^{1/4} 
\left(\frac{10^9\:{\rm K\:cm^{-3}}}{\psc/k}\right)^{3/8}\:{\rm yr},\\
\label{tsfhighmass2}
&\rightarrow & 1.29 \times 10^5 
\left(\frac{m_{*f}}{30\:{M_\odot}}\right)^{1/4} 
\scl^{-3/4}~~~ {\rm yr}.
\end{eqnarray}
For example, with the fiducial values of our parameters, a
$100\:{M_\odot}$ star forming in a clump with $\scl=1$ g cm$^{-2}$ has
a final accretion rate of $1.1\times 10^{-3}\:{M_\odot\:{\rm
yr}^{-1}}$ and a star-formation time of $1.74\times 10^5\:{\rm yr}$.
If the value $\scl\simeq 0.24$ g cm\ee\ inferred for the Orion
Nebula Cluster was appropriate for the molecular gas out of which
the Trapezium stars formed, then the time scale to form
a 30~$M_\odot$ star in Orion was about $3.8\times 10^5$ yr.
The dependence of $\tsf$ on $\krho$ is shown in Figure 1.
For example, had we adopted the value of $\krho=1.8$ found by Mueller et al.
(2002b), the coefficient in equation (\ref{tsfhighmass2}) would
have dropped to $8.9\times 10^4$ yr. 
The star-formation time as a function of final stellar mass is shown
in Figure \ref{fig:tfmf}.  Effects of changing the pressure and using
a different, though still representative, value of $k_\rho$ are also
shown. 

    Our results for the accretion rate and star formation time have
been presented in terms of the surface density of the star-forming
clump.  However, if higher resolution observations are available,
then it is 
convenient to express these results in terms of the measured
density profile, $\nh=\nhref(r/\rref)^{{-k_\rho}}$.  
Note that the resolution must be high enough so that the
data apply to the {\it core} forming an individual star,
and not to the clump in which many stars are forming;
for the fiducial case, this requires measuring the structure
on a scale less than about 0.05~pc$\sim 10^4$~AU (eq. \ref{eq:r2}).
Expressing
the density in terms of the mass, we have
\beq
\nh=\left[\frac{4\pi \muh\nhref^{3/\krho}\rref^3}{(3-\krho)M}\right]
	^{\krho/(3-\krho)},
\eeq
where $\muh=2.34\times 10^{-24}$ g is the mean mass per H.
We now ask, how long will it take the gas in $M$ to form a star
of mass $\ecore M$?  Equation (\ref{tsf}) yields
\begin{eqnarray}
\tsf&=&\frac{1}{\phi_*\krho}
	\left(\frac{3\pi}{8G}\right)^{1/2}
	\left[\frac{(3-\krho)^{6-\krho}}{(\nhref\muh)^{3}}
	\left(\frac{m_*}{4\pi\ecore\rref^3}\right)^\krho\right]
	^{1/2(3-\krho)}\\
&\rightarrow& 5.79\times 10^4
	\left(\frac{10^8\;{\rm cm}^{-3}}{\nhref}\right)
	\left(\frac{1000\;{\rm AU}}{\rref}\right)^{3/2}
	\left(\frac{m_*}{30\;M_\odot}\right)^{1/2}~~~~{\rm yr}.
\end{eqnarray}
Had we chosen $\krho=1.8$, the typical value found by Mueller et al.
(2002b), the result is only slightly different:
\beq
\tsf\rightarrow 8.07\times 10^4
	\left(\frac{10^8\;{\rm cm}^{-3}}{\nhref}\right)^{5/4}
	\left(\frac{1000\;{\rm AU}}{\rref}\right)^{9/4}
	\left(\frac{m_*}{30\;M_\odot}\right)^{3/4}~~~~{\rm yr}.
\eeq
The median value of $\nhref$ at $\rref=10^3$ AU in the Mueller et al.
sample is $2.4\times 10^7$ cm\eee\ (based on the Ossenkopf \& Henning
1994 opacities), which gives $\tsf\simeq 5\times
10^5$ yr for a 30 $M_\odot$ star.  Recall that this choice of
dust opacity led to masses that averaged 3.4 times less than
the virial mass; if the clouds were at their virial masses,
the densities would be  3.4 times higher and the median time
to form a 30 $M_\odot$ star would be $10^5$ yr.

   Recently, Doty et al. (2002) have estimated the timescale
for massive star formation using chemical clocks.  From an
analysis of the source AFGL 2591, they infer that the
chemical
timescale is in the range $7\times 10^3-5\times 10^4$ yr, with
a preference for $3\times 10^4$ yr.  The protostellar models
discussed in \S 6 below suggest that the central star has
a mass of about 13 $M_\odot$. The time scale measured by
Doty et al. refers only to a part of the star formation time,
since the chemistry is strongly affected by the rapidly
rising luminosity. The time scale they infer is in qualitative
agreement with our results.  Unfortunately, our results do
not apply to the the case $\krho=1$ they used to model
their data, since such a model cannot be in a self-gravitating
equilibrium (see Appendix B); as a result, we cannot make a quantitative
comparison.

	Henriksen (private communication) has pointed out that the
$m_{*f}^{1/4}$ dependence of the star-formation timescale was
presented in Henriksen \& Turner (1984) and Henriksen (1986). This
scaling is to be expected, since it follows directly 
from the assumption of hydrostatic equilibrium. In
these models it was assumed that stars form when the turbulence
becomes sub-sonic and that the pressure is a constant in GMCs. 
As we have seen, the former assumption is incorrect for massive stars,
and the
latter assumption leads to a significant over-estimate of 
the star-formation time for massive stars.

	Adams (private communication) has pointed out that 
results similar to ours
are implicit in Adams and Fatuzzo (1996).  By dividing
their estimate for the mass of a protostar by the accretion
rate $c^3/G$, one finds that the star formation times are
about $10^5$ yr and depend only weakly on the mass.  Their
work is based on the assumptions that (1) protostellar accretion
is halted when the rate at which mass falls directly onto
the star, rather than through a disk, is the same as the
mass loss rate in the protostellar outflow; and (2)
that the wind luminosity is a fixed fraction of the stellar
luminosity.  The validity of these assumptions,
particularly for  massive
star formation, remains to be established.

\subsection{Turbulent Core Accretion vs. Competitive Accretion}

An alternative model for massive star formation is the competitive
accretion model, in which massive stars form by accreting gas that was
initially not gravitationally bound to the star.
The accreting stars compete for the mass available in the clump,
and it has been suggested that this process leads to the IMF
(Bonnell et al. 1997; Bonnell et al. 2001).
By contrast,
the turbulent core model is based on the premise that most of the gas that
goes into a star was initially bound.  Since the core is turbulent and
is embedded in a turbulent medium, this is undoubtedly an
oversimplification---at the very least, some of the gas that was bound
to the core at the onset of star formation will become unbound during
the course of the star formation, and similarly some gas that was
initially unbound will become bound. First we estimate the amount of
mass exchange between the core and the clump, and then consider the
rate of accretion directly onto a star once it has formed from its
core.

In general, accretion occurs at a rate given by
\beq
\label{comp}
\dot m_{\rm acc}=\pi \rho_{\rm cl} v_{\rm rel} R_{\rm acc}^2, 
\eeq
where $\rho_{\rm cl}=\phirc\bar{\rho}_{\rm cl}$ is the density of the
ambient clump material, $v_{\rm rel}$ is the relative velocity between
this material and the accreting object, be it core or star, and
$R_{\rm acc}$ is the accretion radius, which is the largest possible
impact parameter of approaching gas that ends up bound to the core or
star. Recall that the factor $\phirc\simeq2.6$ (\S\ref{sec:cores})
allows for the central concentration of clump density
distribution and massive star formation.
We assume
that the relative velocity of the core or star and the ambient medium
is
$\surd 3\sigma_{\rm cl}
=\surd 3 (\phipc/\phirc)^{1/2}\lsrcl$.  The rms velocity dispersion
in the clump, $\lsrcl$, is related to the virial parameter $\alv$ by equation
(\ref{eq:alv}), under the assumption that the clump is spherical.
We then find
\beq
\label{vrel}
v_{\rm rel} = \surd 3 \left(\frac{\phipc}{\phirc}\right)^{1/2}\langle
\sigma_{\rm cl}^2\rangle^{1/2} 
= 4.47 \left(\frac{\phipc}{\phirc}\right)^{1/2} \Sigma^{1/4} \alv^{1/2} \left(\frac{\mcl}{4000\sm}\right)^{1/4}\:{\rm km\:s^{-1}}.
\eeq

\subsubsection{Accretion by the core}

Equation (\ref{tsftffcl}) shows that in typical Galactic
star-forming clumps, the star formation time is comparable to the
free-fall time of the clump, which in turn is comparable to the time
for the core to cross the clump.  Since the surface density of the
core and the clump are similar (see eq. \ref{eq:scscl}) we anticipate
that the core will interact with a mass comparable to its own as the
star forms, or equivalently the rate at which a core interacts with
clump mass should be similar to the rate of star formation.

To make this idea more quantitative,
we estimate $R_{\rm acc}$ for the core by considering particle orbits
that intercept its surface. In this case $R_{\rm acc}=(1+v_{\rm
esc}^2/v_{\rm rel}^2)^{1/2}R_{\rm core}$, where $v_{\rm esc}$ is the escape
velocity from the core,
\begin{eqnarray}
\label{vesc}
v_{\rm esc} & = & 2.27 \left(\frac{m_{*f}}{30\sm}\right)^{1/4}
\ecore^{-1/4} \left(\frac{k_P^2 \phipc \phipb}{A}\right)^{1/8}
\scl^{1/4}\:\:{\rm km\:s^{-1}},\\
 & \rightarrow & 3.00 \left(\frac{m_{*f}}{30\sm}\right)^{1/4} 
\scl^{1/4}\:\:{\rm km\:s^{-1}}.
\end{eqnarray}
The gravitational enhancement of the geometric cross-section is thus
\beq
\label{phigrav}
\phi_{\rm grav}\rightarrow 1+ 0.59 \alv^{-1}
\left(\frac{m_{*f}}{30\sm}\right)^{1/2} 
 \left(\frac{\mcl}{4000\sm}\right)^{-1/2}
\eeq
The enhancement is significant only for massive cores in relatively
low mass clumps.
The rate of accretion is thus
\begin{eqnarray}
\dot{m}_{\rm acc} &=& 2.50\times 10^{-4} \left(\frac{A\phirc 
f_g^2 \alv\phi_{\rm grav}^2}{k_P^2 \ecore^2\phipb}\right)^{1/2} 
\left(\frac{m_{*f}}{30\sm}\right) \left(\frac{\mcl}{4000\sm}\right)
^{-1/4}\scl^{3/4}\smyr,\\
\label{macccore}
 &\rightarrow& 7.9\times 10^{-4} \left(\frac{\phi_{\rm grav}}
{1.6}\right) \left(\frac{m_{*f}}{30\sm}\right) \left(\frac{\mcl}
{4000\sm}\right)^{-1/4}\scl^{3/4}\smyr.
\end{eqnarray}
Thus we see that a massive core will tend to interact with clump
material at a rate that is 
comparable to
the rate at which
it is collapsing or being eroded by star formation,
$\dot{m}_*/\ecore$.  This rate is a factor of several smaller in super
star clusters with $\mcl\sim 10^6\sm$.

	 What is the fate of this swept-up mass?  Gas between the
cores (``intercore gas") is likely to have an Alfven velocity that is
greater than the Keplerian velocity at the surface of the core in view
of its low density; it will therefore not be bound to the core after
being swept up.  It is likely that most of the mass is in cores,
however, and insofar as the core mass distribution follows the IMF,
most of this mass will be in low-mass cores.  When one of these cores
strikes the outer region of a massive core, it is quite likely that it
will strip gas from the massive core, since the relative velocities
are $\gtrsim v_{\rm esc}$. On the other hand, if it strikes the inner
parts of the massive core, it is likely to be absorbed by the larger
core.  Accretion of ambient material onto the massive core will
decrease as soon as a protostellar outflow forms, and it will decrease
further when the protostar becomes massive enough to generate
significant radiation pressure.  Thus, the interaction of the core
with the ambient clump will result in an exchange of material that
most likely causes the massive core to grow, but at a smaller rate
than suggested by equation (\ref{macccore}).  In fact, this growth
may be the mechanism that causes an initially bound, but stable, core
to evolve into an unstable core that collapses to form a star.  This
growth will not be a runaway process since $\rc$ cannot grow
significantly once the expansion wave reaches the edge of the core.

\subsubsection{Bondi-Hoyle accretion by the star}

In the case of competitive accretion, it is the protostar
that accretes gas from the clump.  In this case, the
appropriate accretion radius is given by $R_{\rm acc}=R_{\rm
BH}\simeq 2 G m_* / (v_{\rm rel}^2 + c^2)$ (Bondi \& Hoyle 1944).
We generalize this result by including the effects of magnetic
fields, $c^2=\phi_B\sigma^2$, so that
\beq
R_{\rm BH}=\left(\frac{\phirc/\phipc}{3+\phi_B}\right)
	\frac{2Gm_*}{\lsrcl}.
\eeq
Evaluating this expression in terms of the virial parameter $\alv$,
we obtain
\begin{eqnarray}
R_{\rm BH} & = & \frac{10}{\alv}\left(\frac{\phirc/\phipc}{3+\phi_B}\right)
       \left(\frac{m_*}{\mcl}\right)\rcl\\
 & \rightarrow & 2.89 \times 10^{-3} \alv^{-1} 
\left(\frac{m_*}{10\sm}\right) 
\left(\frac{\mcl}{4000\sm}\right)^{-1/2}\scl^{-1/2}\:\:{\rm pc},
\end{eqnarray}
where we made use of the relation
$\rcl=(\mcl/\pi\scl)^{1/2}=0.516(\mcl/4000\,M_\odot)^{1/2}\scl^{-1/2}$~pc.

	This estimate for the accretion radius is actually an upper limit
for two reasons: First,
it is based on the assumption
that the accreting gas is of uniform density (on scales of $R_{\rm
BH}$) so that it can efficiently dissipate in the wake of the star.
Second, Bondi-Hoyle accretion will cease once the stellar mass is $\gtrsim
10\sm$, since then the main sequence radiation pressure is enough to
counter the star's gravitational attraction, thereby preventing the
focusing of gas streamlines in the wake of the star
(Bonnell et al. 1998).  
We therefore normalize our expressions 
to a stellar mass of 10 $M_\odot$ in this
subsection. 

	 With this result for the accretion radius, we find that
the Bondi-Hoyle accretion rate in a star-forming clump is
\begin{eqnarray}
\dot{m}_{*,{\rm BH}} & = & 1.66\times 10^{-5} \left[\frac{\fg
\phirc^{5/2}}{\alv^{3/2}\phipc^{3/2} (3+\phi_B)^2}\right]
 \left(\frac{m_*}{10\sm}\right)^{2} \left(\frac{\mcl}
{4000\sm}\right)^{-5/4}\scl^{3/4}\smyr,\\
 & \rightarrow & 1.27\times 10^{-6} \alv^{-3/2}
\left(\frac{m_*}{10\sm}\right)^{2} \left(\frac{\mcl}{4000\sm}
\right)^{-5/4}\scl^{3/4}\smyr.
\end{eqnarray}
We see that the Bondi-Hoyle accretion rate (for a $10\sm$ star,
the maximum possible for BH accretion)
is more than two orders of magnitude lower than the turbulent core
accretion rate calculated in \S 4.1. If the clump survives for
$\sim$1~Myr while it forms a star cluster then the maximum fractional change in
a star's mass is $\sim 0.1 (m_*/10\sm) (\mcl/4000\sm)^{-5/4}\scl^{-3/4}$,
so this process will not significantly
alter the stellar mass under typical conditions of massive star
formation.

\begin{figure*}
\epsscale{1.0}
\plotone{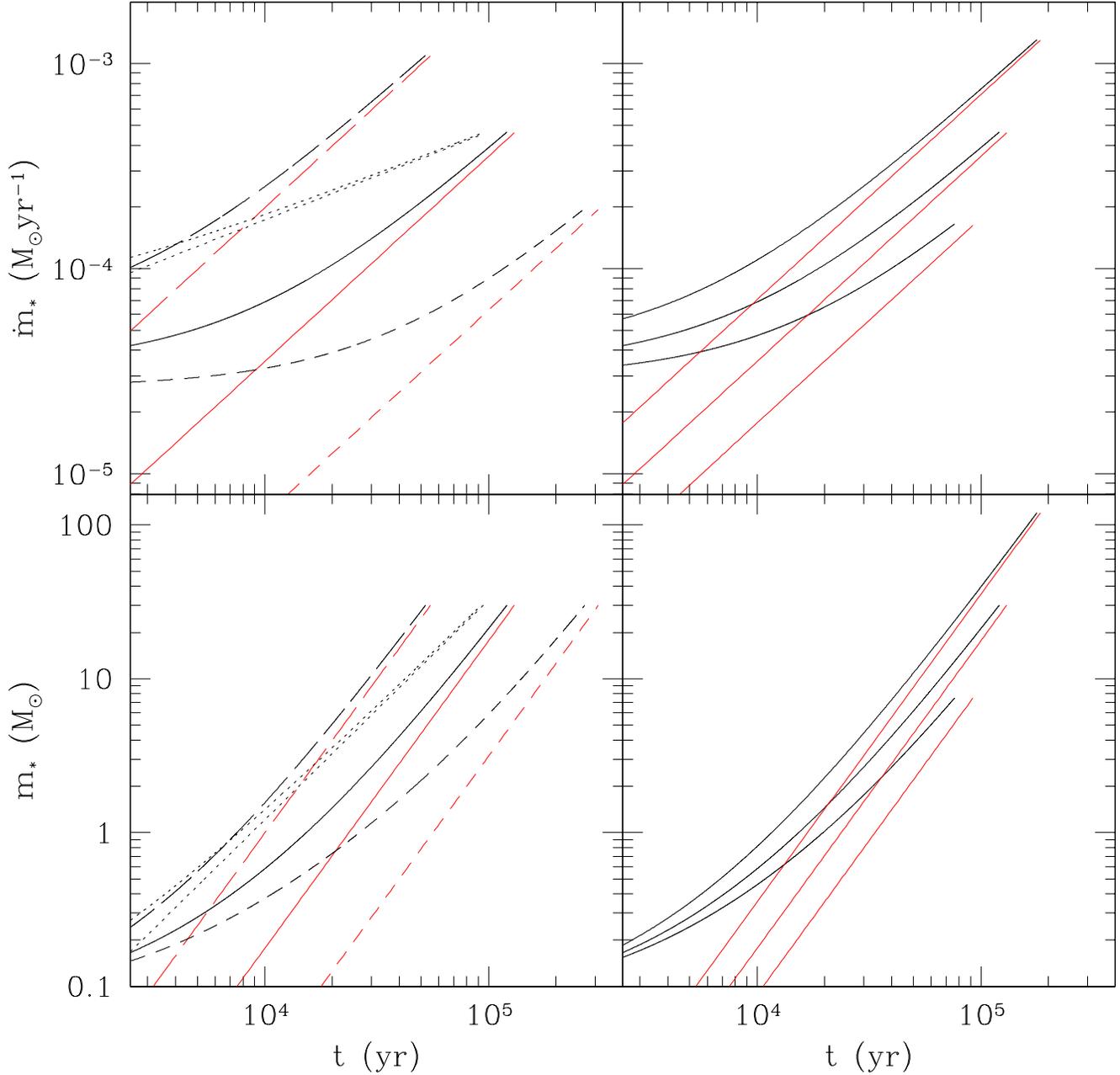}
\caption{Evolution of $\dot{m}_*$ ({\it top panels}) and $m_*$ ({\it
bottom panels}). The straight lines show the purely nonthermal model,
while the curved lines show the two component (thermal [100K] +
nonthermal) core model described in \S\ref{sec:two}. The {\it left
panels} show only $m_{*f}=30\sm$ models. The different cases are
$k_\rho=1.5$ with $\Sigma_{\rm cl}=0.316,1.0,3.16\:{\rm g\:cm^{-2}}$
({\it short-dashed, solid, long-dashed}, respectively), and
$k_\rho=1.75$ with $\Sigma_{\rm cl}=1\:{\rm g\:cm^{-2}}$ ({\it dotted}
line). The {\it right panels} show cases with $m_{*f}=7.5,30,120\sm$
(bottom to top, respectively) for $\Sigma_{\rm cl}=1\:{\rm
g\:cm^{-2}}$ and $k_\rho=1.5$.
\label{fig:mdott}}
\end{figure*}

\begin{figure*}
\epsscale{1.0}
\plotone{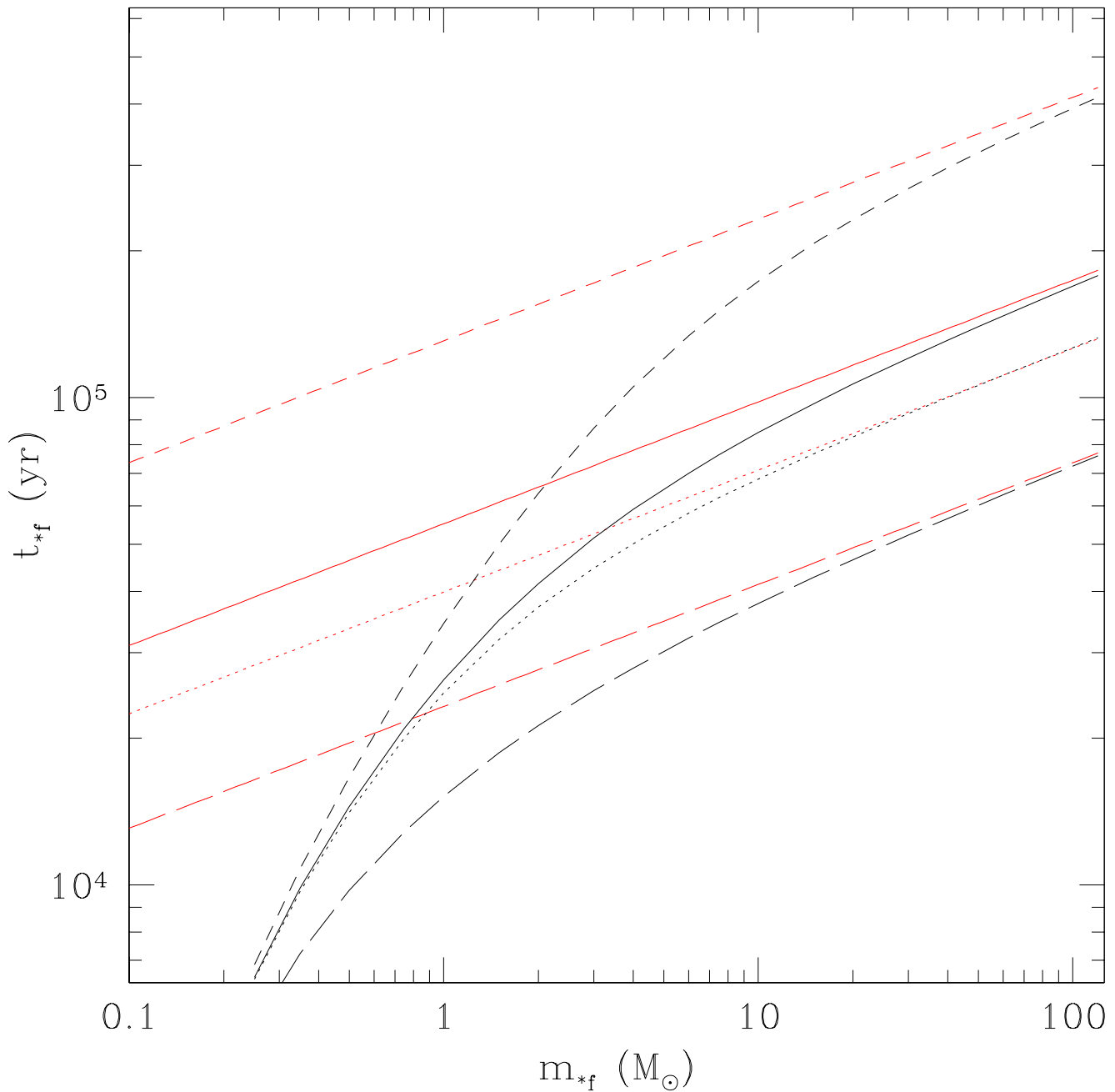}
\caption{Star formation time $t_{*f}$ versus final stellar mass $m_{*f}$ for
accretion from cores with $k_\rho=1.5$ at $\Sigma_{\rm cl}=0.316,1,3.16\:{\rm
g\:cm^{-2}}$ ({\it short-dashed}, {\it solid}, {\it long-dashed}
lines) and with $k_\rho=1.75$ at $\Sigma_{\rm cl}=1\:{\rm g\:cm^{-2}}$ ({\it
dotted} line). As in Figure \ref{fig:mdott}, the straight lines show the purely nonthermal model,
while the curved lines show the two component (thermal [100K] +
nonthermal) core model described in \S\ref{sec:two}.
\label{fig:tfmf}}
\end{figure*}

\section{Two--Component Cores}
\label{sec:two}

Cores in molecular clouds are usually observed to have central regions
dominated by thermal motions and envelopes dominated by nonthermal
motions. 
The presence of a significant thermal pressure component in
the inner regions of cores breaks the self-similarity of the
structure observed on larger scales
in molecular clouds (Williams et al. 2000).
As discussed in \S \ref{S:selfsim}, the nonthermal
pressure component is modeled with $\gamma_p<1$ (Maloney 1988). 
Two components can be modeled as a multipressure polytrope in which
$P=P_{\rm th}+P_{\rm nth}$, where $P_{\rm th}\equiv \rho c_{\rm th}^2$ is
the thermal pressure and $P_{\rm nth}$ is the nonthermal pressure
(McKee \& Holliman 1999). 
To allow for damping of nonthermal motions on small scales,
the gas core can also be modeled as a composite polytrope, in which
the central region is thermally supported and the envelope is
supported by both thermal and nonthermal motions (Curry \& McKee
2000). For analytic modeling, however, the simple ``TNT'' (thermal
plus nonthermal) model introduced by Myers \& Fuller (1992) and
developed by Caselli \& Myers (1995) is convenient. In this model, the
density is assumed to be of the form
\begin{equation}
\label{rhoTNT}
\rho \propto   \left(\frac{r_0}{r}
     \right)^{k_\rho}+\left(\frac{r_0}{r}\right)^2,
\end{equation}
with $1<k_\rho<2$. We assume that $(\rc/r_0)^{2-k_\rho}\gg1$, 
so that the 
first
term
dominates at the surface of the core 
and the model is equivalent to the sum of a
singular polytropic sphere and a singular isothermal sphere:
\begin{equation}
\label{rhoTNT2}
\rho = \rho_s \left(\frac{\rc}{r}\right)^{k_\rho}+
\frac{c_{\rm th}^2}{2\pi G r^2},
\end{equation}
where, by construction, $\rho_s$ is very nearly equal to the density
at the surface; it is therefore given by equation (\ref{eq:rho2}).

   To determine the mass-accretion rate $\dot{m}_*$ and the
star-formation time $t_{*f}$, we need to know 
the value of the free-fall time, $t_{\rm ff}\propto
[\rho(m)]^{-1/2}$. Since the TNT model is just a rough approximation
to the actual density distribution, we can use equation (\ref{eq:m}) to
replace the TNT density distribution $\rho(r)$ with a TNT density
distribution $\rho(M)$,
\begin{eqnarray}
\label{rhoTNTM}
\rho & = & \rho_s \left(\frac{\mcore}{M}\right)^{2(1-j)}+
\frac{2 c_{\rm th}^6}{\pi G^3 M^2},  \\
\label{rhoTNTth}
	& \equiv & \rho_s\left(\frac{\mcore}{M}\right)^{2(1-j)}
     \left[ 1 + \left(\frac{\mth}{M} \right)^{2j}\right],
\end{eqnarray}
where $j$ is defined in equation (\ref{eq:j}).
Here $\mth$ is the mass below which the thermal density distribution 
dominates, and it will be evaluated numerically below.\footnote{
It must be kept in mind that $\mth$ is defined in terms of
the density profile and not the sound speed.
Evaluating $c^2/\cth^2$ with the aid of equation (\ref{eq:c}),
we find
$$
\frac{c^2}{c_{\rm th}^2} = \left(\frac{4}{A k_P^2}\right)^{1/3} 
\left(\frac{M}{\mth}\right)^{2j/3}.
$$
For an isothermal gas ($\gamma_p=1$, $\krho=2$ and $j=0$), 
we have $c^2=\cth^2$ as expected.
However, as $\gamma_p$ decreases below unity 
the value of $c^2$ 
at $M=\mth$ increases
above $\cth^2$.  Although the nonthermal pressure is greater
than the thermal pressure at $\mth$, the two pressure gradients
are comparable there.
}

	  The protostellar accretion rate is
\beq
\mds=\phi_*\;\frac{m_*}{t_{\rm ff}} \propto
	\frac{m_*}{\tffs}\cdot \phi_*
	\left(\frac{\rho}{\rho_s}\right)^{1/2}
\label{eq:mds2}
\eeq
from equation (\ref{eq:dim}).
Let $\phi_{*\rm th}=0.975 \pi \sqrt{3}/[8(1+H_0)
^{1/2}]$ be the
value of $\phi_*$ for the isothermal component, and 
$\phi_{*\rm nth}$ be
the value for the nonthermal component.  
Generalizing equation (\ref{eq:mds2}) with the aid
of equation (\ref{rhoTNTth}) for the density of a two-component
core, we find
\begin{equation}
\label{mdotTNT}
\mds\simeq \dot m_{*f}\left[\left(\frac{m_*}{\msf}\right)^{2j}
     +\left(\frac{\phi_{* \,\rm th}}{\phi_{*\,\rm nth}}\right)^2
     \left(\frac{\ecore\mth}{\msf}\right)^{2j}\right]^{1/2}.
\end{equation}
At early times, when the mass of the protostar is less
than $\ecore\mth$, the accretion rate is constant---it
is given by the isothermal rate (eq. \ref{eq:shu})
corrected by a factor $\ecore/(1+H_0)^{1/2}$ for the star
formation efficiency and the effects of the magnetic field.
Once the mass of the protostar exceeds this characteristic
mass, the accretion rate rapidly approaches the nonthermal rate
evaluated in \S \ref{sec:self}. 
The evolution in time of the accretion rate and instantaneous stellar
mass are shown in Figure \ref{fig:mdott} for two-component cores with
a $T=100\:{\rm K}$ thermal core. The star formation time versus final
stellar mass is shown in Figure \ref{fig:tfmf}.

	We evaluate the mass of the thermal core, $\mth$,
in terms of the Bonnor-Ebert mass, which 
is the maximum mass of a stable isothermal
sphere of gas (Bonnor 1956; Ebert 1955), 
\begin{eqnarray}
\label{mbe}
\mbe =1.182\;\frac{c_{\rm th}^4}{(G^3 \psc)^{1/2}}
& = & 0.0464
\left(\frac{T}{20\:{\rm K}}\right)^2\left(\frac{10^9 
\:{\rm K~cm^{-3}}}{\psc/k} \right)^{1/2}\:{M_\odot},\\
& \rightarrow & 0.0504 \left(\frac{T}{20\:{\rm K}}\right)^2
  \frac{1}{\scl}\:\: M_\odot,
\end{eqnarray}
where the numerical value is 
for a mean particle mass of $2.33 m_{\rm H}$. 
Equations (\ref{eq:rho2}) and (\ref{mbe}) then imply
\begin{eqnarray}
\label{mth}
\mth & = & \mbe \left( \frac{32}{1.182^6\pi^3 A k_P^2} \right)
    ^{(3-k_\rho)/[12(2-k_\rho)]} \left( \frac{\mbe}{M_f} 
    \right)^{(k_\rho-1)/[2(2-k_\rho)]},\\
    & \rightarrow & 1.23\times 10^{-3}\left(\frac{T}{20\;{\rm K}}\right)^3
        \left(\frac{30\ecore\;M_\odot}{\msf}\right)^{1/2}
	\frac{1}{\scl^{3/2}}~~~M_\odot.
\end{eqnarray}
If $M_{\rm core}$ is comparable to $\mbe$, then $\mth\simeq
\mbe$; as $M_{\rm core}$ increases, $\mth$ decreases and becomes a negligible
fraction of the core mass. 
Using equation (\ref{mth}), we can re-express the accretion rate 
as
\beq
\mds\simeq \dot m_{*f}\left[\left(\frac{m_*}{\msf}\right)^
	   {3(2-\krho)/(3-\krho)}
     +\left(\frac{\phi_{* \,\rm th}}{\phi_{*\,\rm nth}}\right)^2
     \left(\frac{32}{1.182^6\pi^3 Ak_P^2}\right)^{1/4}
     \left(\frac{\ecore\mbe}{\msf}\right)^{3/2}\right]^{1/2},
\eeq
which makes the dependence on $\krho$ clearer.
This form of the accretion rate shows 
that the thermal term dominates throughout
the accretion only if $\msf\la \ecore\mbe$.
In the fiducial case, the condition for the thermal term 
to dominate throughout is 
$\msf<1.5\times 10^{-2}(T/20~\rm K)^2
/\scl~M_\odot$; 
even for a temperature of 100 K, this is only a
fraction of a solar mass.  We conclude that the thermal core is
not important in the formation of stars with $\msf\ga 1~M_\odot$ in
typical regions of massive star formation.

	Finally, we note that the fact that the pressure in 
regions of massive star formation is so high means that the
minimum mass of
an isolated star that can form is quite small,
\beq
m_{*\,{\rm min}}\simeq\ecore \mbe.
\label{eq:mmin}
\eeq
If the temperature is of order 
20 K, as in the case of clustered low-mass star formation
(Jijina, Myers, \& Adams 1999) and
the pressure is that of a typical clump that is forming
massive stars ($\psc\sim\bar\pcl\sim 
0.4\times 10^9$ K cm\eee,
corresponding to $\phipc=1$), 
then
this minimum mass is 
slightly
less than 
$0.04 M_\odot$.  
As star formation
proceeds in the clump, the temperature will rise and so will this
typical minimum mass,
$m_{*\,{\rm min}}\propto \mbe\propto T^2$.
On the other hand,
pressure fluctuations in the clump can reduce the value of
$\mbe$ and allow
the formation of stars with lower masses.
The turnover in the IMF at
low masses 
corresponds to the
paucity of cores that form in regions of unusually high pressure
and low temperature,
where $\mbe$ and therefore $m_{*\, \rm min}$ are small.
In the Trapezium, the IMF has a broad maximum between 0.6 $M_\odot$
and the hydrogen burning limit (Muench et al. 2002).
With $\scl=0.24$ g cm\ee\ from Table 1, 
$\phipc=1$ (since the low-mass stars are not 
concentrated in the center of the clump) and $\ecore=0.5$,
equations (\ref{mbe}) and (\ref{eq:mmin})
predict $m_{*\,{\rm min}}\simeq 0.15\; M_\odot$ in Orion, 
which is qualitatively consistent with this observation.

\section{Protostellar Evolution, Luminosities and Comparison to Hot Molecular Cores\label{S:protoevol}}

The properties of protostars depend on their accretion rates
(e.g., Stahler, Shu \& Taam 1980; Stahler 1988; Palla \& Stahler 1992;
Nakano et al. 2000). In particular, the accretion luminosity is
\beq
\label{lacc1}
L_{\rm acc}  =  \frac{f_{\rm acc} G m_*\dot{m}_*}{r_*}=
4685\left(\frac{f_{\rm acc}}{0.5}\right) \left(\frac{m_*}{30\sm} 
\frac{\dot{m}_*}{10^{-4}\smyr} \frac{10{R_\odot}}{r_*}\right)
\:{L_\odot},
\eeq
where $f_{\rm acc}\leq 1$ is a factor that accounts 
for the energy advected into the star or used to drive
protostellar outflows.  If one were able to directly observe
the protostar, it would be possible to distinguish the
protostellar luminosity from that of the accretion disk.
This is generally not possible for massive stars, so here
we shall model the luminosities
of hot molecular cores (HMCs),\footnote{In 
our terminology, some observed hot molecular ``cores'' should properly
be termed hot molecular ``clumps''. The examples chosen here are likely to be true cores.}
which include the contributions of both the protostar and the
accretion disk. The advection of energy into
the star at the accretion shock at the stellar surface
is usually negligible. 
However, the mechanical luminosity of the protostellar outflow
can reduce $f_{\rm acc}$ below unity.  In Paper II, we show
that magneto-centrifugally driven protostellar outflows,
which are powered by the rotational energy of the
star-disk system, typically 
extract about 
half the total available luminosity from
gravitational infall. Thus we set $f_{\rm acc}=0.5$ and note that
$L_{\rm acc}\simeq L_w$. 

	For stars massive enough to be on the main sequence
($m_*\ga 20\,\sm$ in regions of massive star formation---see below),
we can use the zero age main sequence
(ZAMS) mass-radius relation to evaluate
$r_*$. From the solar metallicity models of Schaller et
al. (1992), we find
$r_*=7.27(m_*/30\sm)^{0.55} \, R_\odot$, provided 
$m_*>1.5\sm$.
The accretion luminosity for stars on the main sequence is then
\begin{eqnarray}
\label{lacc2}
L_{\rm acc} & = & 6360\left(\frac{f_{\rm acc}}{0.5}\right) \left(\frac{m_*}
{30\sm}\right)^{0.45} \left(\frac{\dot{m}_*}{10^{-4}\smyr}\right)
\:{L_\odot},\\
\label{lacc3}
 & \rightarrow & 3.0\times 10^4\left(\frac{f_{\rm acc}}{0.5}\right) 
\left(\frac{m_{*f}}{30\sm}\right)^{1.2} \left(\frac{m_*}{m_{*f}}
\right)^{0.95} \Sigma_{\rm cl}^{3/4}\:{L_\odot}.
\end{eqnarray}
The final expression for $L_{\rm acc}$ (eq.[\ref{lacc3}]) uses
the fiducial massive star accretion rate from equation
(\ref{mdotscl}), again for protostars accreting on the ZAMS.
For smaller masses the protostar has a size larger than the main sequence
size, and the accretion luminosity is correspondingly reduced.

In order to calculate the size of the protostar before it reaches the
main sequence one must model protostellar evolution as the star
accretes material and burns deuterium. D-burning provides a source of
energy to support the star in a configuration with a central
temperature $\sim 10^6\:{\rm K}$, so that the star is larger than when
on the main sequence (central temperature of $\sim 10^7\:{\rm
K}$). After a protostar has exhausted its internal store of D, the
luminosity available from this process becomes limited by the rate of
D accretion\footnote{For consistency with the models of Nakano et
al. (1995; 2000) and Palla \& Stahler (1991; 1992) we have adopted an
interstellar abundance of D relative to H of $2.5\times 10^{-5}$
(Bruston et al. 1981; Vidal-Madjar \& Gry 1984). More recent
observations tend to favor somewhat smaller abundances in the range
$\rm D/H\simeq 1.5-2.2\times 10^{-5}$ (Sonneborn et al. 2000; Moos et
al. 2002; Vidal-Madjar \& Ferlet 2002). These results also indicate
that there are spatial variations in the D abundance within the
Galaxy. Our adopted value is at the upper end of this observed range.
Using a value of $\rm D/H=1.6\times 10^{-5}$ reduces $r_*$ by about
30\% for protostellar masses $\gtrsim M_\odot$ (see also Stahler
1988). This raises $L_{\rm acc}$ and thus reduces the estimates of
$m_*$ listed in Table \ref{tab:proto}. However, the effect is very
small ($<10\%$) for $m_*\sim 10 M_\odot$, becomes even smaller at
higher masses as the internal luminosity becomes more dominant, and
disappears completely once the protostar has joined the ZAMS.}:
\beq
\label{ldeut}
L_D \simeq 150 \left(\frac{\rm D/H}{2.5\times 10^{-5}}\right) 
\left(\frac{\dot{m}_*}{10^{-4}\smyr}\right)\:{L_\odot}.
\eeq
Once the star is so massive that this luminosity provides insufficient
support, it starts to contract towards the main sequence. This takes
of order a Kelvin-Helmholtz time. If the star forms a radiative core
during the contraction, shell burning of deuterium temporarily swells
the protostellar radius by about a factor of two (Palla \& Stahler 1992).

Most previous investigations of protostellar evolution 
considered constant accretion rates (e.g. Stahler, Shu \& Taam 1980;
Stahler 1988; Palla \& Stahler 1992; Nakano et al. 2000), although
Behrend \& Maeder (2001) considered evolution with a growing
accretion rate. To consider protostellar evolution with an accretion
rate evolving according to equation (\ref{mdotTNT}), we have developed
a simple model based on that of Nakano et al. (1995; 2000). This model
accounts for the total energy of the protostar as it accretes and
dissociates matter and, if the central temperature $T_c\gtrsim
10^6\:{\rm K}$, burns deuterium. At all times the protostar is assumed
to be a polytropic sphere of index, $n$, so that $\gamma_p=1+1/n$.  We
have modified Nakano et al.'s model to include additional processes,
such as deuterium shell burning and the transition between convective
and radiative structure, and calibrated these modifications against
the more detailed calculations of Stahler (1988) and Palla \& Stahler
(1991; 1992). 

The basic features of our protostellar evolution model, including its
differences compared to the Nakano et al. model, are as follows.  The
initial conditions are set at $m_*=0.1\:{M_\odot}$ and have a radius
that increases with the initial accretion rate (Stahler
1988). Initially the star is contracting and supporting itself by the
release of gravitational energy. At this stage convection is imperfect
so that $n$ is intermediate between $1.5$ and $3$ with a weak
dependence on $\dot{m}_*$, which we calibrate from Palla \& Stahler
(1992). Once the star's central temperature reaches that required for
D core burning, contraction halts and the star becomes fully
convective ($n=1.5$). We set the D core burning temperature at
$1.5\times 10^6\:{\rm K}$ (Stahler 1988). While there is sufficient D 
to burn, the star maintains this central temperature and its radius grows 
linearly with mass. After the protostar's D has been exhausted, D burning 
occurs only at the rate at which it is accreted into the star. 
Inevitably, as
$m_*$ grows, the D luminosity (eq. [\ref{ldeut}]) becomes insufficient
to support the star and contraction continues, raising the central
temperature.

A radiative barrier forms when the luminosity that can be transported
radiatively, $L_{\rm rad}$, rises to equal $L_{\rm D}$. In the Nakano
et al. model $L_{\rm rad}$ is approximated by the main-sequence
luminosity, $L_{\rm ms}$, of a star of the same mass.
We generalize this and assume that a
radiative barrier forms 
when $L_D\leq f_{\rm rad} L_{\rm ms}$,
where for our adopted zero age main sequence (ZAMS) (Schaller et
al. 1992) and by comparison to Palla \& Stahler (1992),
$f_{\rm rad}=0.33$. We approximate the effect of the formation of the
radiative barrier by changing the polytropic index of the whole star
from $n=1.5$ to 3. Also at this point D can no longer penetrate to the
stellar center so D core burning ceases. However, D will still
continue to burn in a shell in regions hotter than $\sim 10^6\:{\rm
K}$. We approximate the onset of shell burning to be simultaneous with
the formation of a radiative barrier (see Palla \& Stahler 1991 for a
more precise description) and we assume it burns at the rate it is
accreted. Shell burning swells the stellar radius by approximately a
factor of two (Palla \& Stahler 1992), which we take to be a constant
in our models, equal to 2.1. In reality the amount of swelling will
decrease as $m_*$ increases, but we find our simple approximation
to be accurate to within $\sim 10\%$ compared to Palla \& Stahler's
results for this stage in the evolution. When $T_c\gtrsim 10^7\:{\rm
K}$, some hydrogen burning reactions begin and cause the star to become
convective once more. The star is rapidly shrinking at this stage, and
once its size becomes equal to the ZAMS radius (Schaller et al. 1992),
we assume the protostar follows this mass-radius relation for the rest of
its growth.

We find our simple one-zone protostellar evolution model matches the
mass-radius relations of the $10^{-5}$, $3\times 10^{-5}$ and
$10^{-4}\:{M_\odot}$ constant accretion rate cases with
photospheric boundary conditions (no advection) of Palla \& Stahler
(1992) to better than about 10\% over the entire evolution to the main
sequence. Furthermore this range of accretion rates is similar to that
predicted by equation (\ref{mdotscl}) for values of $m_*$ relevant to
pre-main-sequence evolution and for densities typical of high-mass
star-forming clumps. The simplicity of our model allows us to apply it in
numerical studies of the formation of clusters of many stars (Tan \&
McKee 2002b).

\begin{figure*}
\epsscale{1.0}
\plotone{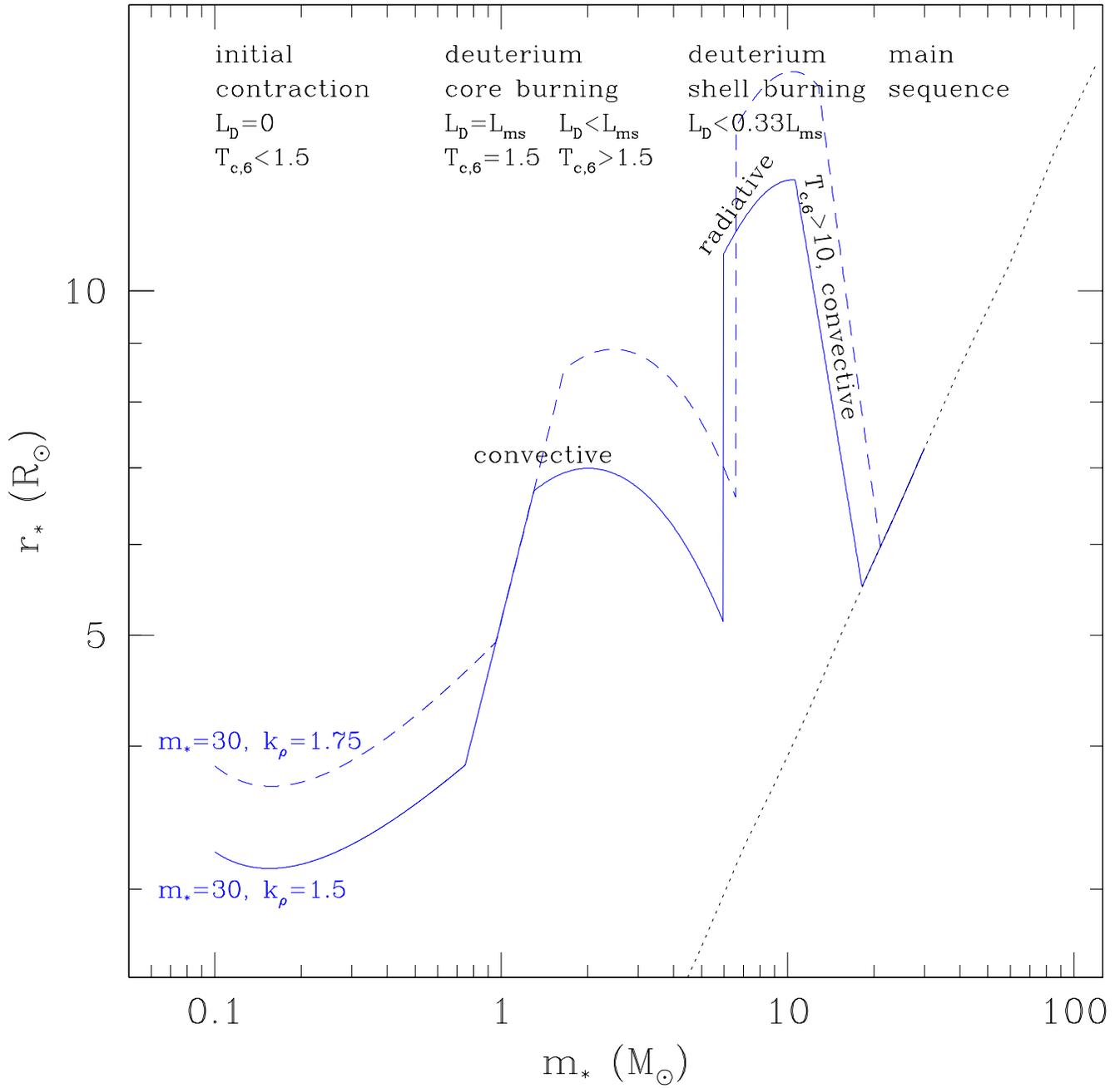}
\caption{Protostellar radius versus $m_*$ for stars with
$m_{*f}=30\:{M_\odot}$ and for $\Sigma_{\rm cl}=1\:{\rm g\:cm^{-2}}$. The
{\it solid} and {\it dashed} lines show the cases with $k_\rho=1.5$
and 1.75, respectively. The {\it dotted} line shows the zero age main
sequence radius from Schaller et al. (1992).
\label{fig:massradplume3}}
\end{figure*}

\begin{figure*}
\epsscale{1.0}
\plotone{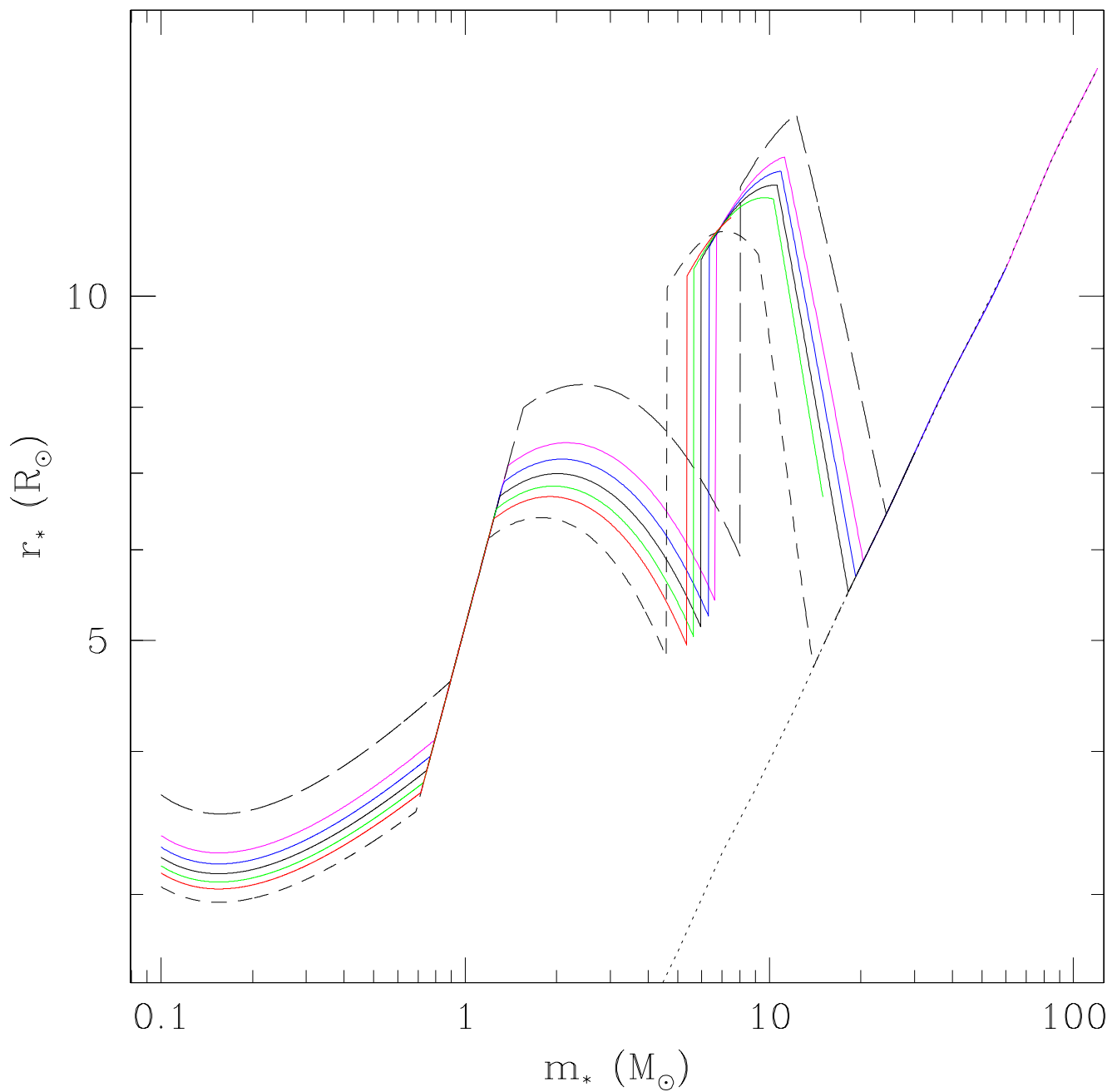}
\caption{Radii of accreting protostars with $k_\rho=1.5$. {\it Solid}
lines show stars of final mass 7.5, 15, 30, 60 and
120~${M_\odot}$ (bottom to top) accreting from cores embedded in a $\Sigma_{\rm cl}=1\:{\rm
g\:cm^{-2}}$ clump, typical of Galactic regions observed by Plume et
al. (1997). The {\it long dashed} and {short dashed} lines show a
30~${M_\odot}$ star forming in a clump with mean pressure 10 and
0.1 times this value, respectively. The {\it dotted} line shows the
zero age main sequence radius from Schaller et al. (1992).
\label{fig:massradplume2}}
\end{figure*}

Figure \ref{fig:massradplume3} shows the evolution of protostellar
radius for $30\:{M_\odot}$ stars forming from cores with
$k_\rho=1.5$ and 1.75 and at pressures equivalent to $\Sigma_{\rm
cl}=1\:{\rm g\:cm^{-2}}$. The accretion rate, here and for all cases
in this paper, is derived using the two-component core accretion model
(equation \ref{mdotTNT}) with $T=100\:{\rm K}$.  As already discussed,
the effect of the thermal core is quite small for the high-pressure
conditions considered here. The accretion rate is dominated by gas supported
by the nonthermal component. The smaller accretion rate of the
$k_\rho=1.5$ case leads to $\sim 25\%$ smaller protostellar radii
during most of the evolution compared to the $k_\rho=1.75$
case. Figure \ref{fig:massradplume2} shows the protostellar radii of
stars of different final masses and illustrates the effect of
increasing and decreasing the core's surface pressure by factors of
10. From equation (\ref{mdotscl}), the accretion rate increases as
$m_{*f}^{3/4}$, $(m_*/m_{*f})^{0.5}$ (for $k_\rho=1.5$) and
$P_s^{3/8}\propto \Sigma_{\rm cl}^{3/4}$. Accreting protostars join
the main sequence at higher masses for higher accretion rates (Palla
\& Stahler 1992). 
For our fiducial case ($\krho=1.5$ and $\scl=1$ g cm\ee), stars
join the main sequence at about 20 $\sm$.  For larger values of
$\krho$ or $\scl$, the accretion rates are higher and
stars join the main sequence at somewhat larger masses
(e.g., for $\msf=30\,\sm$, a star joins the main sequence
at about $24\,\sm$ if either $\krho=1.75$ or $\scl\simeq 3$ g cm\ee).
The maximum protostellar radius also
increases with accretion rate and thus $\Sigma$ of the clump. Again
for $k_\rho=1.5$, $30\:{M_\odot}$ stars reach 10-14~$
R_\odot$ for values of $\scl$ within a factor of 3
of those of the typical Plume et al. clumps.

\begin{figure*}
\epsscale{1.0}
\plotone{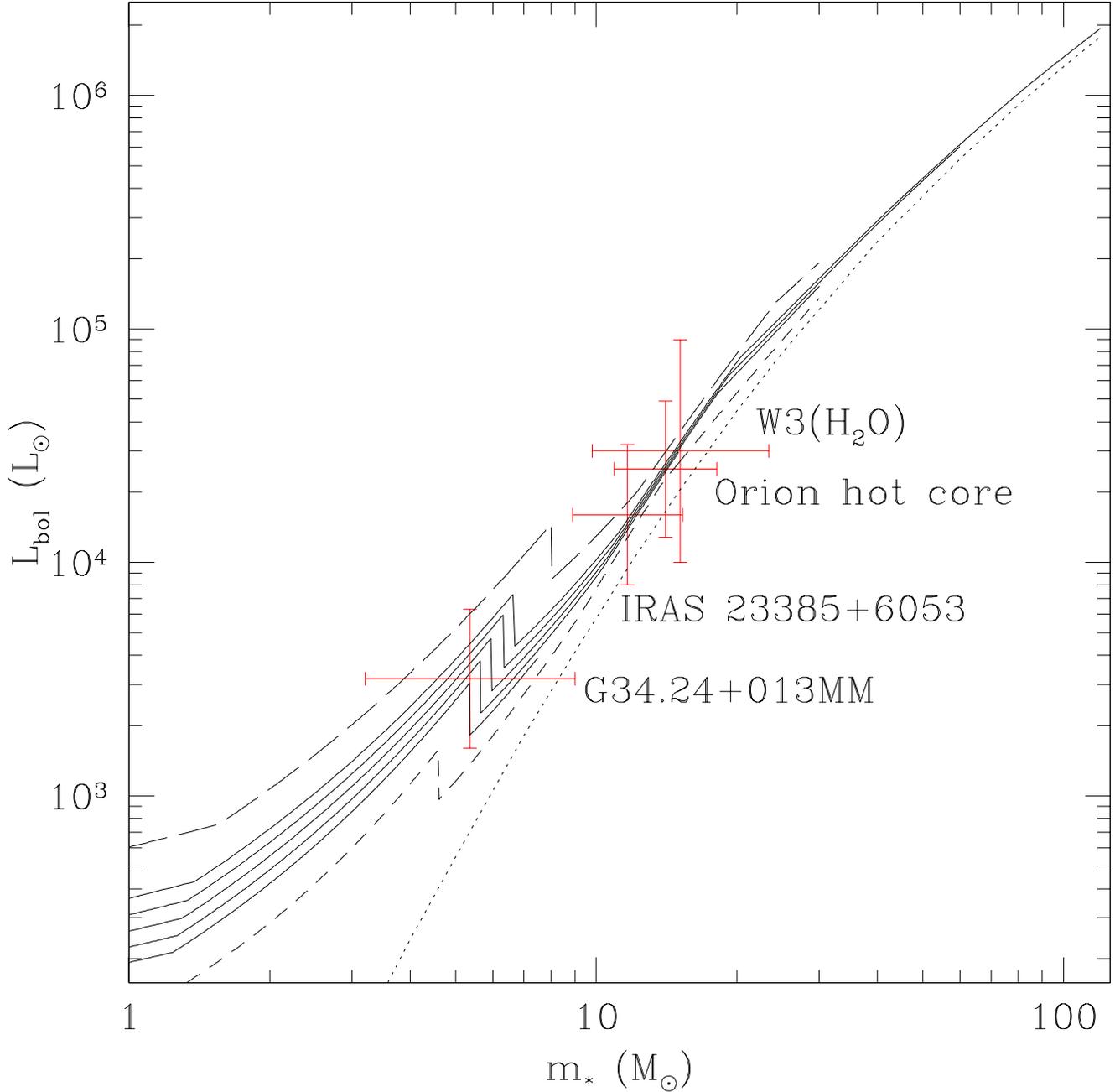}
\caption{Properties of accreting protostars. {\it Solid} lines show
stars of final mass 7.5, 15, 30, 60 and 120~${M_\odot}$ accreting
from cores with $k_\rho=1.5$ embedded in a $\Sigma_{\rm cl}=1\:{\rm
g\:cm^{-2}}$ clump, typical of Galactic regions observed by Plume et
al. (1997). The {\it long dashed} and {short dashed} lines show a
30~${M_\odot}$ star forming in a clump with mean pressure 10 and
0.1 times this value, respectively. The {\it dotted} line shows the
zero age main sequence luminosity from Schaller et al. (1992). The
luminosity step occurring around 5 to 8~${M_\odot}$, depending on
the model, corresponds to the onset of deuterium shell burning, which
swells the protostellar radius by a factor of about two and thus
reduces the accretion luminosity by the same factor. Four observed
HMCs are shown. The vertical error bar illustrates the uncertainty in
their bolometric luminosities. The horizontal error bar shows the
corresponding range of allowed values of $m_*$ for the $\Sigma_{\rm cl}=1\:{\rm
g\:cm^{-2}}$, $k_\rho=1.5$ models (see Table \ref{tab:proto}).
\label{fig:lmplume4}}
\end{figure*}

Our model for the evolution of the protostellar radius allows us to
make predictions for the evolution of the bolometric luminosity of
embedded protostars (Figure \ref{fig:lmplume4}). The bolometric
luminosity is the sum of the internal and accretion luminosities,
$L_{\rm bol}=L_{\rm int}+L_{\rm acc}$. 
We have followed the approximation introduced by Nakano et al. to set
the value of $L_{\rm int}$ approximately equal to $L_{\rm ms}$, the
main sequence luminosity corresponding to the particular value of
$m_*$, and we evaluate $L_{\rm ms}$ from the ZAMS models of Schaller
et al. (1992). Note that before the protostar joins the ZAMS, the
source of $L_{\rm int}$ is gravitational contraction and D burning,
rather than H burning. This approximation is least accurate at small
stellar masses, but in this regime the total luminosity is dominated
by the accretion luminosity.  In fact, it should be noted that in the
fiducial conditions of massive star forming regions, the turbulent
core accretion model predicts quite high accretion luminosities (up to
$\sim 10^3\:{L_\odot}$) when massive protostars are in their early
(low-mass, $\lesssim 3\sm$) stages of formation.

We compare our theoretical tracks with the observed bolometric
luminosities of several HMCs thought to be illuminated internally by
massive protostars. For the purpose of this simple analysis we
consider three values for the initial column density of the clumps in
which these stars started forming, $\Sigma_{\rm cl}=0.32$, 1.0 and $3.2\:{\rm
g\:cm^{-2}}$. This illustrates the effect of a change in the core's
surface pressure by a factor of hundred. We also consider values for
$k_\rho=1.5$ and 1.75. The observed luminosity and its uncertainty
constrain the properties of the protostar. The constraint on $m_*$ is
shown by the extent of the horizontal error bar for four sources in
Figure \ref{fig:lmplume4} for the $\Sigma_{\rm cl}=1\:{\rm g\:cm^{-2}}$,
$k_\rho=1.5$ case. The full list of constraints (on $m_*$, $m_{*f}$
and $\dot{m}_*$) and source properties is shown in Table
\ref{tab:proto}. 

If the pressure (or $\Sigma_{\rm cl}$) is lowered, then the stellar
mass range implied by a given luminosity moves to higher masses, since
the accretion luminosity is reduced and so must be compensated for by
an increase in internal stellar luminosity. Similarly, if the pressure
is raised, then the stellar mass constraints are lowered.
For $L_{\rm bol}\gtrsim {\rm few} \times 10^4\:{L_\odot}$ and for
the conditions considered here, the internal stellar luminosity
(essentially the main-sequence luminosity) tends to dominate that of
accretion.  In these cases the strong dependence of $L_{\rm ms}$ on
$m_*$ means that large pressure uncertainties create only small
uncertainties in $m_*$.

Osorio et al. (1999) have modeled the infrared and submillimeter
spectra of accreting protostars and their surrounding envelopes, and
applied their model to the same sources shown in Figure
\ref{fig:lmplume4} and Table \ref{tab:proto}. We note that
uncertainties in the structure of the gas envelope and the possible
contribution from additional surrounding gas cores or diffuse gas will
affect the observed spectrum. Comparing results, our inferred stellar
masses are similar, except for the lowest luminosity source
G34.24+0.13MM; this discrepancy is a result of 
low value Osorio et al. adopted for the
main-sequence luminosity of a $10\:{M_\odot}$ star.
We systematically favor smaller accretion rates by factors $\sim 2-5$.
Their high accretion
rates of $\sim 10^{-3}\:{M_\odot\:{\rm yr}^{-1}}$ for stars with
$m_*\sim 10\:{M_\odot}$ are difficult to achieve unless the
pressure is increased substantially; for example, if the stars are all
destined to reach $m_{*f}\sim 30\:{M_\odot}$, pressure increases
of a factor $\sim 40$ are required.

\begin{deluxetable}{ccccccc} 
\tablecaption{Properties of Accreting High-Mass Protostars
\label{tab:proto}}
\tablewidth{0pt}
\tablehead{
\colhead{Source [Ref.]\tablenotemark{a}} & \colhead{$L_{\rm bol}$} & \colhead{$k_\rho$} & \colhead{$\Sigma_{\rm cl}$} & \colhead{$m_*$} & \colhead{$m_{*f}$} & \colhead{$\dot{m}_*$}\\
 & \colhead{$({\rm 10^3 L_\odot})$} & & \colhead{$({\rm g\:cm^{-2}})$} & \colhead{$({M_\odot})$} & \colhead{$({M_\odot})$} & \colhead{$({\rm 10^{-4}M_\odot\:yr^{-1}})$}
}
\startdata
W3(${\rm H_2 O}$) [1] & $10-90$ & 1.5 & 1 & $9.8-23.4$ & $\geq 10.5$ & $2.1-5.6$\\
 & & 1.5 & 0.316 & $10.7-25.1$ & $\geq 10.9$ & $0.9-2.5$\\
 & & 1.5 & 3.16 & $6.2-21.1$ & $\geq 9.5$ & $4.6-12.7$\\
 & & 1.75 & 1 & $7.4-23.4$ & $\geq 10.7$ & $2.1-7.7$\\
\hline
Orion hot core [2,3]& $12.6-50$ & 1.5 & 1 & $10.9-18.1$ & $\geq11.3$ & $2.2-5.0$\\
 & & 1.5 & 0.316 & $11.5-19.5$ & $\geq 10.7$ & $1.0-2.2$\\
 & & 1.5 & 3.16 & $6.9-17.1$ & $\geq10.4$ & $4.9-11.4$\\
 & & 1.75 & 1 & $10.5-18.5$ & $\geq 11.6$ & $2.2-7.2$\\
\hline
IRAS 23385+6053 [4]& $8-32$ & 1.5 & 1 & $8.9-15.3$ & $\geq 9.8$ & $2.0-4.6$\\
 & & 1.5 & 0.316 & $10.0-16.3$ & $\geq 10.1$ & $0.9-2.0$\\
 & & 1.5 & 3.16 & $6.6-14.7$ & $\geq 8.5$ & $4.2-10.5$\\
 & & 1.75 & 1 & $6.6-15.8$ & $\geq 10.0$ & $2.0-6.9$\\
\hline
G34.24+0.13MM [5] & $1.6-6.3$ & 1.5 & 1 & $3.2-9.0$ & $\geq 4.2$ & $1.1-3.4$\\
 & & 1.5 & 0.316 & $4.3-9.5$ & $\geq 6.0$ & $0.6-1.5$\\
 & & 1.5 & 3.16 & $2.2-7.6$ & $\geq 3.2$ & $2.0-6.3$\\
 & & 1.75 & 1 & $2.6-9.2$ & $\geq 4.3$ & $1.1-5.2$\\
\enddata
\tablenotetext{a}{\footnotesize References: (1) Wyrowski et al. (1999); (2) Kaufman et al. (1998); (3) Wright et al. (1992); (4) Molinari et al. (1998); (5) Hunter et al. (1998).}

\end{deluxetable}

\section{Conclusions}

We have developed a model (that we term the Turbulent Core Model) for
the formation of massive stars, which is an extension of the classic
paradigm of low-mass star formation (Shu, Adams, \& Lizano 1987) and
is to be contrasted with models involving competitive accretion
(Bonnell et al. 1997, 2001) or stellar collisions (Bonnell et
al. 1998). The principal motivations for the latter models are the
short formation timescales (and correspondingly
high accretion rates) mandated by
observations of short star cluster formation times (Palla \& Stahler
1999) and theoretical considerations of radiation pressure feedback
(Wolfire \& Cassinelli 1987). Such accretion rates are difficult to
justify in the standard picture of isothermal core collapse (Shu
1977). Collapse from cores with nonthermal pressure support can
involve faster accretion rates (Stahler et al. 1980), but there has
been no self-consistent theory for predicting both the normalization
of the expected accretion rates and their evolution. This has led to a
vast range of massive star accretion rates ($10^{-6}-10^{-2}\smyr$)
being considered in the recent literature (e.g. Bernasconi \& Maeder
1996; MP97; Nakano et al. 2000).  Conventional theories of massive
star formation face further problems: massive stars form
preferentially in the centers of stellar clusters (Bonnell \& Davies
1998) where the crowded environment makes it difficult to understand
the existence of massive pre-stellar cores, and the high densities and
pressures lead to a small thermal Jeans mass that is only a fraction
of a solar mass (Bonnell et al. 1998).

The above criticisms, taken together with observational hints that
massive stars may form differently from low-mass stars (central
concentration in star clusters, high degree of equal-mass binarity,
complex morphology and extreme energetics of outflows), have motivated
collisional and competitive accretion models. However, these are also
not without their problems: as noted by Bonnell et al. (1998),
competitive (Bondi-Hoyle) accretion is suppressed for stellar masses
above $\sim 10\sm$ because of radiation pressure feedback, while stellar
collisions require extreme $\gtrsim 10^{8}\:{\rm pc^{-3}}$ stellar
densities, which have never been observed.

The Turbulent Core Model for massive star formation overcomes the
difficulties of the standard accretion scenarios by 
incorporating the effects of the supersonic turbulence and high
pressures observed in
massive
star-forming regions (Plume et al. 1997). The high pressures mean that
cores that become unstable are necessarily very dense and small,
leading to high accretion rates and no over-crowding. The turbulent
and nonthermal nature of cores gives them substructure and thus the
protostar's accretion rate will exhibit fluctuations about the mean.
Note that while clumpiness in the core is an attribute of our model
for massive star formation, it is not a {\it requirement}, as in the
collisional model of Stahler et al (2000). 
The turbulence and
nonthermal support of the clump (protocluster) gas 
determines the IMF for stars above about a solar mass, whereas the
Bonnor-Ebert mass is important in determining the IMF at lower
masses.
While our theory does not aim to predict
the IMF, we note that the empirical core mass function is not too
different from the stellar one (Testi \& Sargent 1998; Motte et
al. 2001), at least up to $\sim 5\sm$, the maximum mass probed by
these observations. More massive cores, with relatively simple,
centrally-concentrated morphologies have been observed, e.g. the Orion
hot core (Wright et al. 1992) and W3($\rm H_2O$) (Wyrowski et al. 1999).
The stellar IMF would then be set by the core mass function, modulated
by $\ecore(m_*)$. In a model in which $\ecore$ is set by the feedback
from bipolar outflows, it is found to have a relatively weak
dependence on stellar mass (Matzner \& McKee 2000; Paper II).

A natural question is why gravitational fragmentation does not
continue down to the thermal Jeans (sub-solar) mass scale inside a
massive core, leading to the formation of a cluster
of low-mass stars instead of a single 
massive star. 
In the self-similar model we have presented, clumps and
cores exhibit density fluctuations on all scales down to the thermal
Jeans mass. We have assumed that clumps 
and cores
are relatively long-lived
(over at least several dynamical times), which implies that most of
the density fluctuations existing at a particular instant are
gravitationally stable. How good is this assumption? The age spread of
pre-main sequence stars in the Orion nebula cluster is of order 1~Myr,
which is about an order of magnitude greater than the expected
free-fall time of a typical Plume et al. (1997) clump. The
approximately spherical morphologies of many clumps observed by
Shirley et al. (2002) suggests that they have existed for at least a
dynamical timescale. If the onset of gravitational instability is a
relatively rare phenomenon in this environment 
(i.e., $\la$ 20\% of the mass of a clump is undergoing gravitational
collapse at any time), as we have assumed,
then it is perhaps not
surprising that gravitational fragmentation in a collapsing core is
unlikely. 
Gravitational fragmentation is further suppressed by the
tidal field of the embedded stars and protostars.

The main results of this paper are the following:

1. The surface densities in observed regions of massive star formation 
   and in star clusters that contain, or did contain, massive stars,
   are typically within a factor 4 of $\scl\sim 1$ g cm\ee.  
   The corresponding mean pressures are
   $\bar P/k\simeq G\scl^2\sim 10^8-10^9$~K~cm\eee, 
   much greater than in
   the diffuse ISM or the typical location in a GMC.

2. Cores have column densities similar to that of the clump in which
   they are embedded.  The mean density of a core significantly
   exceeds that of its natal clump, and the radius is less than the
   the tidal radius.

3. Cores that form massive stars are supersonically turbulent; there
   is no need for the gas to become subsonic in order for star
   formation to occur.

4. The star-formation time is several times the mean free-fall time of
   the core out of which the star forms, but is about equal to that of
   the region in which the core is embedded.

5. The time for a massive star to form in a typical region of
   of massive star formation is about $10^5$~yr,
   with a weak ($m_{*f}^{1/4}$) dependence on stellar mass
   and a somewhat stronger dependence on the surface density of the
   clump in which it is forming ($\scl^{-3/4}$). 
   This
   timescale is short compared to estimated cluster formation times,
   but long compared to the ages of observed supernova remnants, which
   sometimes have been invoked as star formation triggers.

6. The corresponding accretion rate, approaching $10^{-3}\smyr$ for
   the most massive stars, is high enough to overcome the radiation
   pressure due to the luminosity of the star.

7. For the typical case we consider, in which the cores out of which
   the stars form have a density structure $\rho\propto r^{-1.5}$, the
   protostellar accretion rate grows with time as $\dot m_*\propto t$.
   These density structures are consistent with observed clumps and
   cores, while in an appendix we have shown that logatropic models
   are inconsistent.

8. The rate at which a core accretes mass from the ambient clump is
   comparable to the rate at which it processes matter into a star.
   Once the star has formed, subsequent Bondi-Hoyle accretion is 
   negligible, particularly for massive stars.

9. Presenting a calculation of the evolution of the radius of a
   protostar, we determine the protostellar accretion luminosity.
   When the (eventually) massive protostar is still less than a few
   solar masses, this luminosity can be several hundred to a thousand
   solar luminosities. Massive protostars join the main sequence at around
   $20\sm$.

10. Application to observations of the Orion hot core suggests a
   current protostellar mass of between about 11 and 18~$\sm$ and an
   accretion rate of a few$\times 10^{-4}\smyr$. Similar properties
   are estimated for W3($\rm H_2O$), the Turner-Welch object.

The incorporation of feedback, including protostellar outflows,
ionization and radiation pressure, is the subject of Paper II. In
particular the question of when feedback prevents accretion is
addressed. The implications of this work for star cluster formation
will be examined in a future paper (see Tan \& McKee 2002b for an
initial discussion).

\acknowledgments We thank Steve Stahler, Malcolm Walmsley and Richard
Larson for helpful discussions. 
The comments of an anonymous referee improved the clarity of the
paper.
The research of CFM and JCT has been
supported by NSF grant AST-0098365 and by a grant from NASA that
supports the Center for Star Formation Studies. JCT has also received
support via a Spitzer-Cotsen Fellowship from the Department of
Astrophysical Sciences and the Society of Fellows in the Liberal Arts
of Princeton University, and from NASA grant NAG5-10811.

\appendix

\section{Mean Pressure in a Cloud}

	We wish to determine the 
mean pressure in a cloud. 
We assume that the cloud is ellipsoidal,
with radius $R$ normal to the
axis of symmetry and radius $Z$ along the axis. 
The total mass of the cloud is the sum of the
gas mass and the stellar mass, $M=\mg+M_*$. 
Recalling that the pressure is related to the effective
isothermal sound speed $c$ by $P\equiv\rho c^2$, we have
\beq
\bar P\equiv\frac{1}{V}\int P dV
     =\bar\rho\lcr=\left(\frac{3\mg}{4\pi R^2Z}\right)\lcr,
\eeq
where $\bar \rho$ is the volume average of the gas density
and $\langle c^2\rangle\equiv \mg^{-1}\int c^2d\mg$ 
is the mass average of $c^2$.

    The {\it virial mass} of a spherical cloud is defined by
(e.g., Myers \& Goodman 1988)
\beq
\mvir\equiv\frac{5\langle \sigma^2\rangle R}{G},
\label{eq:mvir}
\eeq
where $\sigma$ is the one-dimensional velocity dispersion.
If the pressure in the cloud is entirely due to thermal and
turbulent motions of the gas, then $\sigma=c$; if magnetic
fields contribute, then $\sigma$ differs from $c$, and we
define $\phi_B\equiv \lcr/\lsr\geq 1$.

       The {\it virial parameter} for a spherical cloud is defined as 
(Bertoldi \& McKee 1992),
\beq
\alv\equiv\frac{5\langle \sigma^2\rangle R}{GM}=\frac{\mvir}{M}.
\label{eq:alv}
\eeq
Note that $\alv$ is defined in terms of the total mass $M$;
the gas mass is a fraction $\fg$ of the total, $\mg=\fg M$.

	These definitions are readily extended to non-spherical
clouds (Bertoldi \& McKee 1992).  For an individual cloud,
$R$ is replaced by $R_{\rm obs}=(R_{\rm Max}R_{\rm min})^{1/2}$,
where $R_{\rm max}$ and $R_{\rm min}$ are the maximum and minimum
observed radii in the plane of the sky.  
For an ensemble of clouds, the value of $R_{\rm obs}$
averaged over orientation is denoted by $R_m$, and the virial
mass and virial parameter are defined in terms of $R_m$.
In particular, axisymmetric ellipsoidal 
clouds with a volume $4\pi R^2Z/3$ have 
\beq
\phig\equiv\frac{R_m^3}{R^2Z}
\eeq
in the range $1\leq \phig\leq 1.41$
for $0.3<Z/R<3$.  For both prolate clouds with $Z=2R$ and
oblate clouds with $Z=0.5R$, we have $\phig=1.13$.
Recent observations of star-forming clumps do not show strong evidence
for asphericity (Shirley et al. 2002). For our numerical estimates
we consider the spherical case with $\phig=1$.

      Combining these results, we find that the mean pressure is
related to the surface density of the cloud
(including any embedded stars), 
$\Sigma\equiv M/\pi
R_m^2$, by
\beq
\bar P\equiv \left(\frac{3\pi \fg \phi_B\phig\alv}{20}\right)G\Sigma^2.
\label{eq:pbara}
\eeq
Note that this is an identity, since it follows directly from the
definitions of the various quantities involved.  The physics enters
primarily in the determination of the virial parameter $\alv$,
which measures the effect of gravity, and in the
magnetic parameter $\phi_B$.
Below, we estimate that $\alv\simeq 1$ and $\phi_B\simeq 2.8$.
We define the factor in parentheses in equation (\ref{eq:pbara}) as $\phipb$.
If we set $\fg=2/3$ as in the text and 
$\phig=1$, then we have
\beq
\bar P\equiv \phipb G\Sigma^2\simeq 0.88 G\Sigma^2.
\eeq

\subsection{The Virial Parameter $\alv$}
\label{sec:alv}

	Existing data suggest that $\alv\simeq 1$ for gravitationally
bound molecular clouds and cores.  Solomon et al. (1987) studied GMCs
in the first Galactic quadrant and found that the virial masses for
the clouds they studied were close to those inferred from gamma ray
observations.  Such observations determine the CO to H$_2$ conversion
factor, $X\equiv N(H_2)/W_{\rm CO}$.  The most recent determinations
of $X$ from EGRET data give $X=1.8\times 10^{20}$ cm$^{-2}$/K km
s$^{-1}$ in the local ISM (Dame, Hartmann, \& Thaddeus 2001) and
$X=1.56\times 10^{20}$ cm$^{-2}$/K km s$^{-1}$ in the Galactic plane
(Hunter et al. 1997).  Determining $X$ from virial masses, Solomon et
al. found $X_{\rm vir}=2.2\times 10^{20} /M_6^{0.235}$
cm$^{-2}$/K km s$^{-1}$, where $M_6$ is
the cloud mass in units of $10^6 M_\odot$. [Their masses have been
increased by a factor 1.31 to allow for a reduction in the distance to
the Galactic Center from 10 to 8.5 kpc (a factor 0.85), for an
increase in the mass due to the extrapolation to zero intensity
recommended by Solomon and Rivolo (1989; a factor 1.4), and a
difference in the definition of $\mvir$ (a factor 1.1).] Note that
$X_{\rm vir}/X=N_{\rm vir}/N=M_{\rm vir}/M=\alv$. For a typical cloud mass of
$10^6~M_\odot$ in the Galactic plane, 
this gives $\alv=X_{\rm vir}/X({\rm Galactic~plane})
=1.4$.  Note that the gamma-ray determination of $X$ is necessarily
biased toward nearby clouds, which have an average mass somewhat less
than $10^6~M_\odot$, which could raise $\alv$ somewhat.  In any case,
Solomon et al. estimated their cloud masses based on the assumption
that $\alv=1.1$; if it is in fact larger than this, then their masses
should be reduced by a factor $1.1/\alv$.  Dame et al. (1986) surveyed
the brightest CO clouds in the first Galactic quadrant and determined
cloud masses with $X=1.6\times 10^{20}$
cm$^{-2}$/K km s$^{-1}$.  Reducing their masses by
$0.85^2$ and their radii by 0.85 to adjust to our adopted distance to
the Galactic Center, we find that the mean virial parameter in their
cloud sample is $\alv=1.3$, with a standard deviation of 0.76.  Their
results do not give a dependence of $\alv$ on cloud mass as found by
Solomon et al. (1987), although it must be kept in mind that the Dame
et al. sample is considerably smaller.

	Both bound and unbound structures exist within GMCs.
Bertoldi \& McKee (1992) found that most $^{13}$CO clumps are
gravitationally unbound; however, the most massive ones, which
were also the ones that showed evidence for star formation,
appeared bound.  Tachihara, Mizuno, \& Fukui (2000) have studied
the C$^{18}$O cores in Ophiuchus and shown that they are generally
bound.  In the northern region, the value of $\alv$ calculated
from the mean properties of the clouds (and including the
thermal contribution of H and He to $\sigma^2$) is
2.2 (excluding one core with $\mvir=12M$), whereas in the main
body of Ophiuchus it is 0.9.  In Taurus, Onishi et al. (1996)
found 40 C$^{18}$O cores with $\alv=0.9$ as calculated
from the mean properties (and with the same correction
to $\alv$).  The fraction of these cores that have
embedded young stellar objects is
16\%, 33\%, and 48\% in the northern
region, main body of $\rho$ Oph, and Taurus, respectively.
The two regions with the most active star formation thus have
$\alv\sim 1$.  However, these data refer only to regions
of low-mass star formation.  There are no data from which
one can directly determine the virial parameter for massive-star forming
regions such as those studied by Plume et al. (1997). 

	The value of $\alv$ can also be estimated from theory.
Let ${\cal T}=\frac32 M\langle\sigma^2\rangle$ 
be the total kinetic energy (including thermal energy)
and ${\cal W}\equiv \frac 35 a GM^2/R$ be the gravitational
energy; the parameter $a$ is the ratio of the gravitational
energy to that of a uniform sphere.
The virial parameter is then
$\alv=2a{\cal T}/|{\cal W}|$. 
Including the effects of 
the surface pressure
$P_s$ and the magnetic fields, the virial theorem implies
\beq 
\alv=\frac{a}{(1-P_s/\bar P)}\left(1-\frac{\cal M}{|{\cal W}|}
	\right),
\eeq
where ${\cal M}$ is the magnetic energy (Bertoldi \& McKee 1992).
The value of ${\cal M}/|{\cal W}|$ is quite uncertain, however.

In the text, we model the cores as singular polytropic
spheres (SPSs).  If we adopt the same model for the clump in 
which the cores are embedded, we have
\beq
\alv=\frac{\lsr}{\lcr}\left(\frac{5\lcr R}{GM}\right)=
	\frac{\alpha_{\rm SPS}}{\phi_B},
\label{eq:alv2}
\eeq
where 
\beq
\alpha_{\rm SPS}= \frac 52\left(\frac{4-3\gamma_p}{6-5\gamma_p}\right) 
	\left(\frac{2-\gamma_p}{\gamma_p}\right)
\label{eq:alvsps}
\eeq
is the virial parameter of an SPS (see MP96 or McKee \& Holliman
1999).  As discussed in \S 3, observed clumps have density
distributions characterized by values of $k_\rho$ ranging from 1 to 2, 
with 1.5 being typical.  The value of $\gamma_p$ corresponding
to $k_\rho=1.5$ is 2/3, which gives $\alpha_{\rm SPS}=15/4$.
Below, we estimate $\phi_B=2.8$ so that $\alv=1.34$,
which is comparable to the observed values.

\subsection{The Magnetic Parameter $\phi_B$}

	The effective sound speed $c^2\equiv P/\rho$ includes
the contributions of both the gas pressure and the magnetic
pressure,
\beq
c^2=\sigma^2+\frac{B^2}{8\pi\rho}+\frac{\delta B^2}{24\pi\rho},
\eeq
where $\sigma$ is the 1-D velocity dispersion, which can be inferred
from observation.  We have distinguished between the pressure
associated with the background field $B$ (pressure = energy density)
and that associated with a turbulent field $\delta B$ [assumed
to have a random orientation so that pressure
=(1/3) energy density].  
Taking the mass average of this relation gives
\beq
\lcr=\lsr\left(1+\frac 32 \frac{E_B}{E_K}+\frac 12 \frac{E_{\delta B}}
	{E_K}\right),
\label{eq:c2}
\eeq
where $E_B=\int(B^2/8\pi)dV$ is the background field energy,
$E_K=\frac 32 \fg M\lsr$ is the total kinetic energy of the gas, 
and $E_{\delta B}$ is
the energy of the turbulent field.  The second term can be
rewritten in terms of an average Alfven Mach number $m_A$, where
\beq
m_A^2\equiv \frac{3\lsr}{\langle v_A^2\rangle}=\frac{E_K}{E_B}.  
\eeq

	We now show that the last term in equation
(\ref{eq:c2}) is about $1/3$.
Fluctuations in the magnetic field are associated
with motions perpendicular to the field, and, for small
amplitudes and negligible ambipolar diffusion, these motions
are in equipartition with the fluctuating field (Zweibel \& McKee 1995):
$(1/2)\rho v_\perp^2=\delta B^2/8\pi$.  There are also motions
along the field; assuming that the velocity field is approximately
isotropic ($v_\perp^2=2\sigma^2$), we find $(2/3)E_K=E_{\delta B}$.
Thus, for isotropic turbulence the last term is $1/3$.
One can show that the same result is obtained if one considers the fluctuations
to be MHD waves in a cold plasma
and evaluates the wave pressure using the results
of Dewar (1970).  This result is in reasonably good agreement
with the simulations of supersonic
MHD turbulence by
Stone, Ostriker, and Gammie (1998), who found
$E_{\delta B}/2E_K=(0.28, 0.30)$ for
$\beta\equiv 8\pi P_{\rm th}/B^2=(0.01, 0.1)$, respectively.
We therefore adopt
$E_{\delta B}/2E_K=0.3$ for low $\beta$, supersonic turbulence,
which is the type that appears to be characteristic in molecular
clouds.  The effect of the magnetic field is then described by
\beq
\phi_B\equiv \frac{\lcr}{\lsr}=
	1+\frac 32 \frac{E_B}{E_K}+\frac{E_{\delta B}}{2E_K}
	=1.3+\frac{3}{2m_A^2}
\label{eq:fb}
\eeq
from equation (\ref{eq:c2}).  This equation does not apply
to high-$\beta$ plasmas, which have $\phi_B\simeq 1$.

	The magnitude of the magnetic field in massive-star forming
clumps is unknown. The most complete
survey of magnetic field measurements in star-forming regions is
that of Crutcher (1999).  While this survey primarily deals with
regions of low-mass star formation, it does include regions whose
density is comparable to, and even greater than, that in the Plume
et al. clumps.  For regions in which the magnetic field 
could be measured, the median value of the Alfven Mach number is
1.0.  The magnetic field factor 
is then $\phi_B=1.3+1.5=2.8$.  

\section{The Failure of Logatropes}

	   Turbulent regions have velocity dispersions that increase
with scale.  Larson (1981) found that molecular clouds
have line widths that increase with size at a rate intermediate
between that of Kolmogorov turbulence ($\lsr^{1/2}\propto r^{1/3}$)
and that of Burgers' turbulence, which is appropriate for a system of
shocks ($\lsr^{1/2}\propto r^{1/2}$).  Subsequent observations have
shown that GMCs (Solomon et al 1987; Heyer \& Schloerb 1997)
and low-mass cores (e.g., Caselli \& Myers 1995) have line widths
that increase approximately as $r^{1/2}$.  Larson (1981) also found
that the density tends to fall off as $1/r$ (i.e., $k_\rho=1$), so 
that in these regions there is a relation
between the velocity dispersion and density, $\lsr^{1/2}\propto
\rho^{-1/2}$. As discussed in the text, regions of high-mass
star formation typically have $k_\rho\simeq 1.5$, so they usually
do not satisfy this line-width density relation.  Nonetheless,
inferences drawn about the properties of turbulence in low-mass
cores and GMCs would have implications for high-mass cores as well.

	Before continuing, we note that Larson's relations 
primarily refer to properties of an ensemble of clouds, rather
than to the structure of individual clouds.  The relation
$n\propto 1/r$ then means that the column densities of different
clouds are all about the same, to within an order of magnitude.
(This can be understood in terms of the conditions needed for
the stability of molecular clouds--see Elmegreen 1989; McKee 1999.)
If this relation applied within individual clouds, and if
the line-width size relation $\lsr^{1/2}\propto r^{1/2}$ also
applied, then the turbulent pressure $\rho\lsr$ would be
{\it constant} and would not exert a force.  It is for this
reason that the density that can be supported
by a self-similar cloud
in hydrostatic equilibrium (eq. \ref{eq:rho}) vanishes for
$\krho=1$.

	Lizano \& Shu (1989) pointed out that the line-width density
relation $\lsr^{1/2}\propto
\rho^{-1/2}$ 
follows if one assumes that (1) the turbulent pressure
is proportional to the logarithm of the density, $P_{\rm turb}\propto
\ln\rho$, since then the signal speed 
$(dP/d\rho)^{1/2}$ varies as $\rho^{-1/2}$, and
(2) the velocity dispersion
$\sigma$ varies as the signal speed $(dP/d\rho)^{1/2}$.  
Including the thermal pressure of an isothermal gas, they
wrote the total pressure as
\beq
P=\rho\cth^2+K\ln(\rho/\rho_0),
\label{eq:plizano}
\eeq
where $K$ is a numerical constant and
$\rho_0$ is a reference density.  Gehman et al. (1996) 
termed this a ``logatropic'' equation
of state.
Note that logatropes are not self-similar, and therefore
avoid the ``no-force" problem that afflicts self-similar polytropes
with $\krho=1$.

	MP96 pointed out a fundamental problem with this formulation:
Since the central velocity dispersion is observed to be thermal, 
it is necessary to set $\rho_0=\rho_c$, which Gehman et al. (1996)
did.  However,
if one then sets $K\gg 1$, as both Lizano \& Shu (1989) and Gehman et
al. (1996) did, then the pressure becomes negative when the density
drops only slightly from its central value.  For example, if $K=10$,
the pressure is already negative for $\rho=0.9\rho_c$.

	To overcome this problem, MP96 proposed an alternate form of
the logatropic equation of state,
\beq
P=P_c[1+A_{\rm MP}\ln(\rho/\rho_c)],
\eeq
where $A_{\rm MP}$ is a numerical constant that they estimate is about
0.18 in observed molecular clouds.  (We have added a subscript ``MP''
to the parameter $A$ defined by McLaughlin \& Pudritz in order to
distinguish it from the parameter defined in the text. 
We emphasize that our model in the main text is based on the polytropic
results of McLaughlin \& Pudritz, not the logatropic ones.)
They present a
thorough study of the structure and stability of spherical clouds with
this equation of state.  In particular they show that for $A_{\rm
MP}\ll 1$ and at large $r$, the density approaches the form
\beq
\rho=\rho_c(r_1/r),
\eeq
where
\beq
r_1\equiv \left(\frac{A_{\rm MP}\sigma_c^2}{2\pi G\rho_c}\right)^{1/2};
\eeq
indeed, the density is quite close to this value
for $r\ga 1.5 r_1$.  Gravitationally bound clouds have radii $R\gg r_1$.
MP96 focussed on clouds near the brink of instability, which have 
$R/r_1=\exp(1/A_{\rm MP}-1/4)\rightarrow 200$, where the numerical
evaluation is for $A_{\rm MP}=0.18$.  The collapse of logatropic
spheres was studied analytically by MP97 and numerically by Reid,
Pudritz, \& Wadsley (2002).

	Unfortunately, this form of the logatropic equation of state
is also fundamentally flawed: {\it It satisfies the observed line-width
size relation only when the cloud is gravitationally unbound;}
furthermore, {\it it leads to a ``droop'' in the line-width size
relation for clouds that are bound.}
To see this, we evaluate the properties of the MP-logatrope in the case
$r\gg r_1$, so that $\rho\propto 1/r$:
\begin{eqnarray}
M & = & 2\pi\rho_c r_1 R^2,
\label{eq:mmp}\\
\frac{\lsr}{\sigma_c^2} & = & \frac{1}{M\sigma_c^2}\int P dV=\frac{2R}{3r_1}
	\left[1+\frac 13 A_{\rm MP}-A_{\rm MP}\ln\left(\frac{R}{r_1}\right)\right],
\label{eq:smp}\\
\alv & \equiv & \frac{5\lsr R}{GM}
	=\left(\frac{10}{3A}\right)\left[1+\frac 13 A_{\rm MP}-A_{\rm MP}\ln
	\left(\frac{R}{r_1}\right)\right],
\label{eq:alvmp}
\end{eqnarray}
where the virial parameter $\alv$ is discussed in \S \ref{sec:alv}.
Since the effects of magnetic fields have not been taken into
account here, this value of the virial parameter corresponds to
$\alpha_{\rm non}$ in the notation of MP96.  If magnetic fields are
important, then MP96 assume that $\alpha_{\rm non}$ is still given
by the rightmost expression in equation (\ref{eq:alvmp}), while
the observed value (which they denote $\alpha_{\rm mag}$) is reduced.  In
terms of the parameter $\phi_B$ introduced 
in Appendix A, we have $\alpha_{\rm
mag}=\alpha_{\rm non}/\phi_B$.

	Evaluating the exponent in the line-width size relation for
MP-logatropes, we find
\beq
\frac{d\ln\lsr}{d\ln r}=1-\frac{10}{3\alv}.
\eeq
This result immediately reveals the two flaws described above.
First, these logatropes obey the observed line-width size relation
$\lsr\propto r$ only when the virial parameter $\alv\gg 10/3$,
corresponding to an observed virial parameter $\alpha_{\rm mag}\gg 1$;
i.e., the
clouds are unbound.  
Second, when clouds are strongly bound (i.e., nearly critical), the
line width actually {\it decreases} with size.  
This effect is clearly apparent in Figure 2 of MP96.
In particular, 
for a critical cloud, we have $\alv=35/18$
(eq. \ref{eq:alvmp}; MP96) so that the slope of the line-width size
relation is -5/7.  Such a ``droop'' in the line-width size relation
has never been observed.  
For example, the observations of four GMCs 
by Heyer \& Schloerb (1997) show that in each case the
velocity dispersion rises approximately as $r^{1/2}$ over the
entire range of their observations. (However, it should be noted
that observations of high-mass star-forming clumps are not yet
sufficiently detailed to rule out the possibility of a droop
in the line-width size relation in these objects.)
We conclude that, where tested by observation, this form of the logatropic
equation of state also fails.



\end{document}